\RequirePackage{fix-cm}
\documentclass[smallextended]{svjour3}       
\smartqed  
\usepackage{amsmath}
\usepackage{graphicx,epsfig,wrapfig}
\usepackage{makeidx}
\usepackage{amsfonts}
\usepackage{amssymb}
\usepackage{enumitem}
\usepackage{amsmath}
\usepackage{dsfont}
\usepackage{mathrsfs}
\usepackage{slashed}
\usepackage{epstopdf}
\usepackage{bm}
\usepackage{color}
\usepackage[normalem]{ulem}
\usepackage{simplewick}

\newcommand{\ud}{\mathrm{d}}

\newcommand{\uTr}{\mathrm{Tr}}

\newcommand{\ucal}{\mathcal}

\newcommand{\uvec}[1]{\boldsymbol{#1}}

\newcommand{\uind}[2]{^{#1}_{\phantom{#1}#2}}

\begin{document}

\title{The relativistic center of mass in field theory with spin}

\author{C\'edric Lorc\'e}

\institute{Centre de Physique Th\'eorique, \'Ecole polytechnique, 
	CNRS, Universit\'e Paris-Saclay, F-91128 Palaiseau, France\\\email{cedric.lorce@polytechnique.edu}}

\date{\today}

\maketitle

\begin{abstract}
In order to unravel the origin of the nucleon spin, one has to study in detail the question of orbital angular momentum, and in particular the reference point about which it is defined. With this in mind, we review the concept of relativistic center of mass, generalize the discussion to the case of asymmetric energy-momentum tensors, and establish the link with the light-front formalism. We find that the $p$-wave in the Dirac plane-wave solutions arises from a relativistic quantum-mechanical effect which forces the canonical reference point to depend on the observer. This explains why longitudinal spin is much simpler to study than transverse spin. It is also the reason behind the observation of induced shifts and distortions in the parton distributions defined within the light-front formalism.
\end{abstract}


\section{Introduction}

One usually expects to recover ordinary Quantum Mechanics from Quantum Field Theory (QFT) by considering the non-relativistic limit. While position is considered to be an observable in Quantum Mechanics, it is demoted to a mere parameter in QFT. How can one then recover a position operator in the non-relativistic limit?

The absence of a position operator in QFT is usually explained by the fact that a particle with mass $m$ can at best be localized over distances of the order of the associated reduced Compton wavelength $\bar\lambda_C=\hbar/mc$. If one tries to localize a particle over shorter distances, energies larger than $mc^2$ have to be involved, with the risk of creating additional particles in the system. Consequently, the concept of position for a single particle has to be abandoned in QFT. 

The energy-momentum tensor (EMT) of a system can however always be used to define the concept of center of inertia\footnote{By inertia we mean inertial mass, i.e. the system's resistance to being accelerated by a force.}. For a non-relativistic particle, it should coincide with the actual position of the particle. In QFT, the center of inertia may be considered as the closest we can get to a definition of position observable. Systems with non-zero spin are however delicate, since in that case it seems impossible to find a Lorentz covariant definition for the center of inertia leading to commuting coordinates~\cite{Pryce:1948pf}.

In QFT, the questions involving position are often simply ignored or avoided. Contrary to the electroweak sector of the Standard Model, the elementary constituents of Quantum Chromodynamics never show up in the spectrum of asymptotic states. In other words, experiments can only detect bound states of quarks and gluons. Understanding the origin of the nucleon spin becomes therefore a natural fundamental question to investigate. In particular, since orbital angular momentum is defined relative to a reference point, the problem of nucleon localization cannot be avoided. Despite intense efforts invested in the decomposition of the nucleon spin~\cite{Leader:2013jra}, we feel that this problem has so far been insufficiently treated in the literature. 

The present work aims at filling this gap. We start with a reminder of the definition of various contributions to the Poincar\'e generators in Section~\ref{sec2}. Then we review in Sections~\ref{sec3} and~\ref{sec4} the concept of center of inertia assuming as usual that the EMT is symmetric. We generalize in Section~\ref{sec5} this discussion to the more general case of systems with intrinsic angular momentum naturally characterized by an asymmetric EMT, and establish in Section~\ref{sec6} the connection with the light-front formalism widely used to investigate the internal structure of the nucleon. To keep the presentation simple, we work at the level of classical fields in flat spacetime, but the quantum version proceeds analogously\footnote{Typically, forthcoming results can be taken as applying to the quantum theory, provided simple technical adjustments are made. For example, Noether currents and charges become operator-valued distributions acting on a Hilbert space, and classical products $A B$ are replaced by symmetric products $\frac{1}{2}(AB+BA)$ of the corresponding operators.}, as explained in detail in~\cite{Pryce:1948pf,Moller:1949bis,Born:1935ap,Fleming:1965}. We will also use for convenience the natural units $\hbar=c=1$.

\section{Poincar\'e generators}\label{sec2}

The Noether currents associated with the Poincar\'e invariance of a relativistic theory consist in $T^{\mu\nu}(x)$ for spacetime translations, and $M^{\mu\alpha\beta}(x)=-M^{\mu\beta\alpha}(x)$ for Lorentz transformations. At the classical level, they are mere functionals of the classical fields depending on the space-time coordinates $x^\mu$, and are understood to be evaluated for a solution of the classical field equations.

As a consequence of Noether's theorem, these currents are conserved
\begin{equation}
\partial_\mu T^{\mu\nu}(x)=0,\qquad\partial_\mu M^{\mu\alpha\beta}(x)=0,
\end{equation}
and transform as rank-2 and 3 Lorentz tensors, respectively. The 10 generators of Poincar\'e transformations are obtained by integrating the current densities over a spacelike hypersurface
\begin{equation}
P^\mu\equiv\int\ud^3x\, T^{0\mu}(x),\qquad J^{\alpha\beta}\equiv\int\ud^3x\, M^{0\alpha\beta}(x).
\end{equation}
These Noether charges are interpreted as the total four-momentum and generalized angular momentum (AM) of the system. Assuming as usual that surface terms at spatial infinity do not contribute, one can show that these generators are time-independent and transform as rank-1 and 2 Lorentz tensors, respectively.

\subsection{External and internal parts}

The generalized AM $J^{\alpha\beta}$ is defined relative to the origin $O$ of the coordinate system. Like in Newtonian mechanics, one can also consider the generalized AM defined relative to another reference point (or pivot). Denoting by $X^\mu(x^0)$ the function describing the worldline of a generic reference point parametrized by the time coordinate $x^0$ in the Lorentz frame $\ucal S$, the generalized AM 
\begin{equation}\label{extint}
J^{\alpha\beta}=L^{\alpha\beta}_X(x^0)+S^{\alpha\beta}_X(x^0)
\end{equation}
can be decomposed into an external part which depends explicitly on the coordinates of the reference point
\begin{equation}\label{extdef}
L^{\alpha\beta}_X(x^0)\equiv X^\alpha(x^0) P^\beta-X^\beta(x^0) P^\alpha,
\end{equation}
and an internal part
\begin{equation}\label{intdef}
S^{\alpha\beta}_X(x^0)\equiv J^{\alpha\beta}-L^{\alpha\beta}_X(x^0)
\end{equation}
which represents the generalized AM about the new reference point. When the reference point coincides with the origin, one naturally gets $J^{\alpha\beta}=S^{\alpha\beta}_O$. While we keep here $X$ totally general, we will see later that particular choices for the reference point beside the origin, like e.g. the center of inertia, play a particular role.

Setting $\beta=0$ in Eq.~\eqref{intdef} allows us to express the coordinates of the reference point as\footnote{We restrict ourselves to the case of massive systems for which the evaluation of $P^0$ leads to a stricly positive number. In the quantum theory, $1/P^0$ has to be understood as the inverse energy operator $(P^0)^{-1}$ whose eigenvalue when acting on an energy eigenstate coincides with the inverse of the energy eigenvalue.}
\begin{equation}
X^\mu(x^0)=x^0\,\frac{P^\mu}{P^0}+\frac{J^{\mu 0}-S^{\mu 0}_X(x^0)}{P^0},
\end{equation}
where we used the relation $X^0(x^0)=x^0$ resulting from our choice of parametrization for the reference worldline in terms of the time coordinate $x^0$. 

When we consider the time derivative of this expression, we obtain\footnote{To get the four-velocity, one has to multiply this expression by the Lorentz boost factor $\gamma=P^0/M$, where $M$ is the mass of the system.}
\begin{equation}
\dot X^\mu(x^0)=\frac{P^\mu}{P^0}-\frac{\dot S^{\mu 0}_X(x^0)}{P^0},
\end{equation}
where we denoted the derivative of a function of a single variable by a dot. This shows that whenever the velocity of the reference point and the momentum are not parallel $\dot X^\mu(x^0)\not\propto P^\mu$, there is a transfer of generalized AM between external and internal parts $\dot S^{\mu 0}_X(x^0)=-\dot L^{\mu 0}_X(x^0)\neq 0$. 

If one changes the reference point from $X^\mu(x^0)$ to $\widetilde X^\mu(x^0)=X^\mu(x^0)+Q^\mu(x^0)$, the corresponding generalized internal AM are related as follows
\begin{equation}\label{tildeS}
S^{\alpha\beta}_X(x^0)=S^{\alpha\beta}_{\widetilde X}(x^0)+Q^\alpha(x^0)P^\beta-Q^\beta(x^0)P^\alpha,
\end{equation} 
and the shift $Q^\mu(x^0)$ can be expressed as
\begin{equation}\label{Qshift}
Q^\mu(x^0)=\frac{S^{\mu 0}_X(x^0)-S^{\mu 0}_{\widetilde X}(x^0)}{P^0}.
\end{equation}
We naturally have $Q^0(x^0)=0$ since $X^\mu(x^0)$ and $\widetilde X^\mu(x^0)$ represent two simultaneous positions in $\ucal S$.

\subsection{Boost and rotation generators}\label{KJ}

Just like the 6 independent components of the electromagnetic tensor $F^{\mu\nu}(x)=-F^{\nu\mu}(x)$ can be expressed in some Lorentz frame $\ucal S$ in terms of an electric vector $F^{0i}(x)=-E^i(x)$ and a magnetic vector $F^{ij}(x)=-\epsilon^{ijk}B^k(x)$, the 6 independent components of the generalized AM tensor can be expressed in terms of two spatial vectors 
\begin{equation}
J^{0i}=K^i,\qquad J^{ij}=\epsilon^{ijk}J^k,
\end{equation}
which generate boosts and rotations of the system, respectively. Under boosts, the two vectors $\uvec K$ and $\uvec J$ get mixed with each other, indicating that their definition depends on the observer.

A more covariant definition of boost and rotation generators requires the introduction of an object that specifies the observer independently of the frame. Choosing the spatial axes to be parallel in all inertial frames, any observer can be identified by a timelike unit four-vector\footnote{More explicitly, this unit four-vector is the timelike vierbein whose components are $u^\mu=\Lambda\uind{\mu}{0}$, where $\Lambda$ is the Lorentz transformation relating $\ucal S_u$ to $\ucal S$.} $u^\mu$ (taken to be constant in flat spacetime) which represents its four-velocity relative to some Lorentz frame $\ucal S$. The generalized AM tensor can then be written as 
\begin{equation}\label{Jdec}
J^{\alpha\beta}=-K^\alpha_u u^\beta+K^\beta_u u^\alpha-\epsilon^{\alpha\beta\mu\nu}J_{u\mu} u_\nu,
\end{equation}
where we used the convention $\epsilon_{0123}=+1$ and defined the four-vectors\footnote{One might be worried about the change in the spatial integration for the generators defined relative to $\ucal S_u$. However, it follows from Noether's theorem that the Poincar\'e generators transform as ordinary Lorentz tensors, and therefore do not depend on how the spatial integrations are performed.}
\begin{equation}
K^\mu_u\equiv -J^{\mu\nu}u_\nu\qquad\text{and}\qquad J^\mu_u\equiv\tfrac{1}{2}\,\epsilon^{\mu\alpha\beta\lambda}J_{\alpha\beta}u_\lambda.
\end{equation}
These two four-vectors are orthogonal to $u^\mu$ and are hence spacelike. In the observer's frame $\ucal S_u$, the four-velocity reduces to its canonical form $u^\mu=(1,\uvec 0)$ and we get
\begin{equation}
K^\mu_u\overset{\ucal S_u}{=}(0,\uvec K),\qquad J^\mu_u\overset{\ucal S_u}{=}(0,\uvec J).
\end{equation}
This shows that the four-vectors $K^\mu_u$ and $J^\mu_u$ represent in a covariant way the boost and rotation generators as defined by an observer with four-velocity $u^\mu$ relative to the Lorentz frame $\ucal S$. 

Under a change of observer, the covariant boost and rotation generators get mixed with each other
\begin{align}
K^\mu_v&=K^\mu_u(u\cdot v)-u^\mu(K\cdot v)+\epsilon^{\mu\nu\alpha\beta}v_\nu J_\alpha u_\beta,\\
J^\mu_v&=J^\mu_u(u\cdot v)-u^\mu(J\cdot v)-\epsilon^{\mu\nu\alpha\beta}v_\nu K_\alpha u_\beta.
\end{align}
In particular, we find in the $\ucal S_u$ frame
\begin{align}
&&K^0_v&=\uvec K\cdot\uvec v,&\uvec K_v&=\uvec K v^0+(\uvec v\times\uvec J),&\\
&&J^0_v&=\uvec J\cdot\uvec v,&\uvec J_v&=\uvec J v^0-(\uvec v\times\uvec K).&~\label{JLT}
\end{align}

Choosing to parametrize the reference worldline $X^\mu(\tau_u)$ in terms of the Lorentz-invariant expression for the observer's time coordinate $\tau_u=X^\mu(\tau_u)u_\mu$, the covariant boost generators 
\begin{equation}
K^\mu_u=K^\mu_{L_Xu}(\tau_u)+K^\mu_{S_Xu}(\tau_u)
\end{equation}
can be further decomposed into external and internal parts
\begin{align}
K^\mu_{L_Xu}(\tau_u)&\equiv -L^{\mu\nu}_X(\tau_u)u_\nu=-X^\mu(\tau_u)\,(P\cdot u)+\tau_u\,P^\mu,\\
K^\mu_{S_Xu}(\tau_u)&\equiv -S^{\mu\nu}_X(\tau_u)u_\nu,\label{KSXu}
\end{align}
and we can express the coordinates of the reference point as
\begin{equation}\label{refpoint}
X^\mu(\tau_u)=\tau_u\,\frac{P^\mu}{P\cdot u}+\frac{\left[J^{\mu\nu}-S^{\mu\nu}_X(\tau_u)\right]u_\nu}{P\cdot u}.
\end{equation}
Similarly, for the covariant rotation generators we have
\begin{equation}\label{JLS}
J^\mu_u=L^\mu_{Xu}(\tau_u)+S^\mu_{Xu}(\tau_u)
\end{equation}
with
\begin{align}
L^\mu_{Xu}(\tau_u)&\equiv \tfrac{1}{2}\,\epsilon^{\mu\alpha\beta\lambda}L_{X\alpha\beta}(\tau_u)u_\lambda=\epsilon^{\mu\alpha\beta\lambda}X_\alpha(\tau_u) P_\beta u_\lambda,\label{LX}\\
S^\mu_{Xu}(\tau_u)&\equiv \tfrac{1}{2}\,\epsilon^{\mu\alpha\beta\lambda}S_{X\alpha\beta}(\tau_u)u_\lambda.
\end{align}

\subsection{Center-of-mass frame}

Among all the Lorentz frames, the center-of-mass (CM) frame $\ucal S_\star$ defined as the frame where the system is at rest, is special because its four-velocity relative to some Lorentz frame $\ucal S$ is expressed directly in terms of the translation generators $u^\mu_\star=P^\mu /M$ with $M=\sqrt{P^2}$. Physical quantities defined in $\ucal S_\star$ are proper to the system in the sense that they can be expressed in a covariant way using only the Poincar\'e generators. For example, the Lorentz invariant quantity $M=P\cdot u_\star\overset{\ucal S_\star}{=}P^0$ coincides in the CM frame with the energy or inertia of the system. In other words, $M$ represents the proper inertia of the system, explaining why it is usually called the (invariant) mass of the system in the literature.

Using $u^\mu_\star=P^\mu/M$ in Eq.~\eqref{LX} shows that the proper AM is purely internal $J^\mu_\star=S^\mu_\star$, and hence does not depend on the choice of the reference point. It is called the spin of the system, and is proportional to the Pauli-Luba\'nski pseudo-vector $W^\mu=M S^\mu_\star$ defined as
\begin{equation}
W^\mu\equiv \tfrac{1}{2}\,\epsilon^{\mu\alpha\beta\lambda}J_{\alpha\beta}P_\lambda.
\end{equation} 
Moreover, by a suitable choice of the reference point, we can make the proper boost generators purely external. This reference point, called the center of mass of the system, is readily obtained from Eq.~\eqref{refpoint} by setting Eq.~\eqref{KSXu} to zero
\begin{equation}\label{CMpos}
R^\mu_\star(\tau)\equiv\tau\,\frac{P^\mu}{M}-\frac{K^\mu_\star}{M},
\end{equation}
where $\tau=R^\mu_\star P_\mu/M$ is the proper time of the system. The time-dependent term cancels out in the expression for the external part of the generalized AM~\eqref{extdef}, and so the center of mass defines a time-independent decomposition of the generalized AM tensor into external and internal parts. In particular, the proper boost generators $K^\mu_\star=-MR^\mu_\star(0)$ can be interpreted as giving (up to a sign) the mass dipole moment of the system at the initial proper time $\tau=0$.

\subsection{Composite systems}

Composite systems may consist of several types of constituents. For example, hadrons are bound states made of quarks and gluons. In this case, one can naturally split the total currents into partial currents associated with the different constituent types
\begin{equation}
T^{\mu\nu}(x)=\sum_aT^{\mu\nu}_a(x),\qquad M^{\mu\alpha\beta}(x)=\sum_a M^{\mu\alpha\beta}_a(x).
\end{equation}
Unlike the total currents, the partial ones are in general not conserved, and so the corresponding charges usually depend on time
\begin{equation}
P^\mu_a(x^0)\equiv\int\ud^3x\, T^{0\mu}_a(x),\qquad J^{\alpha\beta}_a(x^0)\equiv\int\ud^3x\, M^{0\alpha\beta}_a(x).
\end{equation}

The partial generalized AM can naturally be decomposed into external and internal parts
\begin{equation}
J^{\alpha\beta}_a(x^0)=L^{\alpha\beta}_{Xa}(x^0)+S^{\alpha\beta}_{Xa}(x^0),
\end{equation}
where
\begin{align}
L^{\alpha\beta}_{Xa}(x^0)&\equiv X^\alpha(x^0)P^\beta_a(x^0)-X^\beta(x^0)P^\alpha_a(x^0),\\
S^{\alpha\beta}_{Xa}(x^0)&\equiv J^{\alpha\beta}_a(x^0)-L^{\alpha\beta}_{Xa}(x^0).
\end{align}
These satisfy
\begin{equation}
L^{\alpha\beta}_X(x^0)=\sum_a L^{\alpha\beta}_{Xa}(x^0),\qquad S^{\alpha\beta}_X(x^0)=\sum_a S^{\alpha\beta}_{Xa}(x^0),
\end{equation} 
and can be interpreted as the contributions from constituent type $a$ to the external and internal parts of the generalized total AM.

The partial generalized AM can also be written in terms of partial covariant generators
\begin{equation}
J^{\alpha\beta}_a(x^0)=-K^\alpha_{a,u}(x^0) u^\beta+K^\beta_{a,u}(x^0) u^\alpha-\epsilon^{\alpha\beta\mu\nu}J_{a,u\mu}(x^0) u_\nu,
\end{equation}
where
\begin{equation}
K^\mu_{a,u}(x^0)\equiv -J^{\mu\nu}_a(x^0)u_\nu,\qquad J^\mu_{a,u}(x^0)\equiv\tfrac{1}{2}\,\epsilon^{\mu\alpha\beta\lambda}J_{a\alpha\beta}(x^0)u_\lambda
\end{equation}
satisfy 
\begin{equation}
K^\mu_u=\sum_aK^\mu_{a,u}(x^0),\qquad J^\mu_u=\sum_aJ^\mu_{a,u}(x^0).
\end{equation} 
These can then be interpreted as the contribution from constituent type $a$ to the covariant generators of boosts and rotations.

\section{Center of inertia}\label{sec3}

A natural choice for the reference point is the center of inertia, which is the relativistic generalization of the familiar concept of center of mass appearing in Newtonian Mechanics. Since the inertial mass is given by the energy in a relativistic theory, the center of inertia is determined by the EMT.

The energy dipole moment of the system in some Lorentz frame $\ucal S$ is given by
\begin{equation}
D^\mu(x^0)=\int\ud^3x\,x^\mu T^{00}(x).
\end{equation}
The center of inertia is defined as the point where the inertia of the whole system $P^0=\int\ud^3x\,T^{00}(x)$ can be concentrated without changing the energy dipole moment. Its position is then given by~\cite{Fokker:1929,Pryce:1948pf,Moller:1949,Moller:1949bis}
\begin{equation}\label{CMdef}
R^\mu(x^0)=\frac{D^\mu(x^0)}{P^0}.
\end{equation}
In particular, one finds that $R^0(x^0)=x^0$ which means that the component $R^0$ represents the time $x^0$ at which the center of inertia is determined.

In general, the center of inertia is not at rest in $\ucal S$. Assuming as usual that surface terms at spatial infinity do not contribute, the velocity of the center of inertia can be expressed as
\begin{equation}\label{CMvel}
\dot R^\mu(x^0)=\frac{\ucal J^\mu(x^0)}{P^0},
\end{equation}
where $\ucal J^\mu(x^0)=\int\ud^3x\,T^{\mu 0}(x)$ is the energy current. Unlike the four-momentum $P^\mu=\int\ud^3x\, T^{0\mu}(x)$, the energy current can in general be time-dependent. This means that the center of inertia does not necessarily move along a straight line with constant velocity.

\subsection{Generalized orbital angular momentum}

From the EMT, one can also define an orbital (or convective) tensor
\begin{equation}\label{convdef}
M^{\mu\alpha\beta}_\text{orb}(x)\equiv x^\alpha T^{\mu\beta}(x)-x^\beta T^{\mu\alpha}(x),
\end{equation}
whose four-divergence is given by
\begin{equation}\label{convdiv}
\partial_\mu M^{\mu\alpha\beta}_\text{orb}(x)=T^{[\alpha\beta]}(x)
\end{equation}
with $T^{[\alpha\beta]}=T^{\alpha\beta}-T^{\beta\alpha}$. The corresponding charge
\begin{equation}\label{gOAM}
M^{\alpha\beta}_\text{orb}(x^0)=\int\ud^3x\, M^{0\alpha\beta}_\text{orb}(x)=\int\ud^3x\left[x^\alpha T^{0\beta}(x)-x^\beta T^{0\alpha}(x)\right]
\end{equation}
will be called the generalized orbital angular momentum (OAM) since its purely spatial components represent the total OAM of the system
\begin{equation}
M^{ij}_\text{orb}(x^0)=\int\ud^3x\left[x^i T^{0j}(x)-x^j T^{0i}(x)\right].
\end{equation}
For the other components, we find
\begin{equation}\label{boostconv}
M^{\mu 0}_\text{orb}(x^0)=R^\mu(x^0) P^0-x^0 P^\mu,
\end{equation}
so that the position of the center of inertia can be expressed as
\begin{equation}\label{CMR}
R^\mu(x^0)=x^0\,\frac{P^\mu}{P^0}+\frac{M^{\mu 0}_\text{orb}(x^0)}{P^0}
\end{equation}
and the velocity as
\begin{equation}\label{CMRV}
\dot R^\mu(x^0)=\frac{P^\mu}{P^0}+\frac{\dot M^{\mu 0}_\text{orb}(x^0)}{P^0}.
\end{equation}
The same expression can be obtained from Eq.~\eqref{CMvel} by writing the energy current as $\ucal J^\mu(x^0)=P^\mu+\int\ud^3x\,T^{[\mu 0]}(x)$ and using Eq.~\eqref{convdiv}. Note that unlike~\eqref{CMvel}, the expression for the velocity in Eq.~\eqref{CMRV} does not rely on the assumption that surface terms vanish. 

\subsection{Alternative definition}

Let $X^\mu(x^0)$ denote the position of some reference point at the time $x^0$. Similarly to Eq.~\eqref{intdef}, the internal generalized OAM can be defined as
\begin{equation}
\begin{aligned}
\ell^{\alpha\beta}_X(x^0)&\equiv M^{\alpha\beta}_\text{orb}(x^0)-X^\alpha(x^0)P^\beta+X^\beta(x^0)P^\alpha\\
&=\int\ud^3x\left[(x^\alpha-X^\alpha(x^0))\, T^{0\beta}(x)-(x^\beta-X^\beta(x^0))\, T^{0\alpha}(x)\right].
\end{aligned}
\end{equation}
In particular, $\ell^{\mu 0}_X$ represents the energy dipole moment about the reference point
\begin{equation}
\ell^{\mu 0}_X(x^0)=\int\ud^3x\,(x^\mu-X^\mu(x^0))\, T^{00}(x),
\end{equation}
and $\ell^{\mu 0}_X/P^0$ gives the shift of the center of inertia relative to the reference point
\begin{equation}
\frac{\ell^{\mu 0}_X(x^0)}{P^0}=R^\mu(x^0)-X^\mu(x^0).
\end{equation}
This means that the center of inertia can alternatively be defined by the condition
\begin{equation}
\ell^{\mu 0}_R(x^0)=0.
\end{equation}
Indeed, by construction the center of inertia is the reference point about which the energy dipole moment vanishes. In other words, it is the point for which orbital boost generators are purely external, as already expressed by Eq.~\eqref{boostconv}. In practice we will omit the label $R$ whenever the center of inertia is chosen as reference point.

\subsection{Centroids}

\begin{figure}[t!]
	\centering
	\includegraphics[width=.65\textwidth]{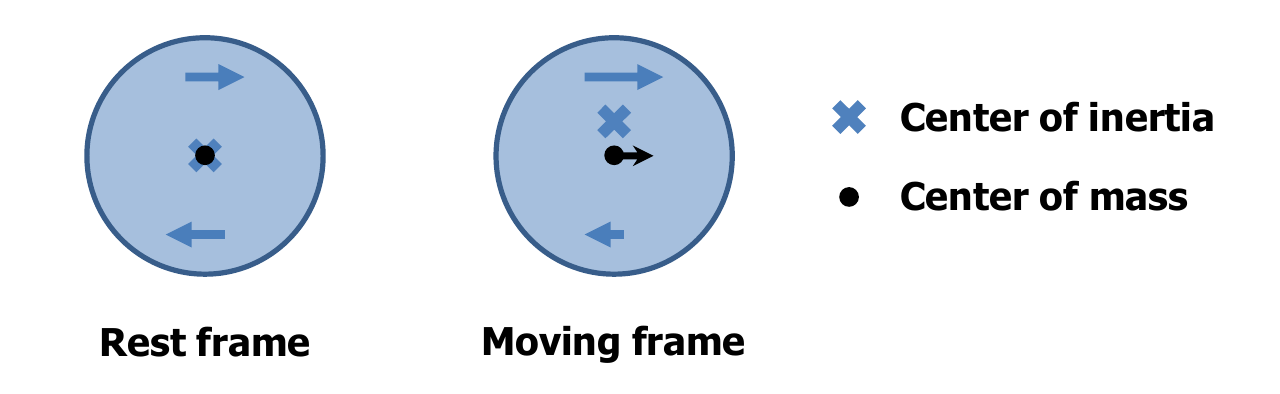}
\caption{\footnotesize{For a spinning homogeneous sphere at rest, center of mass and center of inertia coincide with each other. When the sphere is moving in a direction orthogonal to its spin, the center of inertia moves away from the center of mass. The representation is purely schematic and does not take into account Lorentz contraction factors.}}
		\label{fig1}
\end{figure}

Is is clear from the definition~\eqref{CMdef} that the position of the center of inertia does not transform as a Lorentz four-vector. This is also clearly illustrated by M\o ller's famous example~\cite{Moller:1949,Moller:1949bis}, see Fig.~\ref{fig1}. Consider an homogeneous sphere rotating about some axis in the CM frame $\ucal S_\star$. By symmetry, the center of inertia necessarily lies on the rotation axis. Consider now another frame $\ucal S$ moving with constant velocity in a direction orthogonal to the rotation axis. From the perspective of $\ucal S$, symmetrical points with respect to the rotation axis do not move with the same speed anymore, and hence are attributed different inertias. As a result, the center of inertia determined by $\ucal S$ is shifted in a direction orthogonal to both the rotation axis and the relative velocity between $\ucal S_\star$ and $\ucal S$. This means that the actual center of inertia, i.e. the representative point and not only its coordinates, depends on the observer. It cannot be in general identified with a physical point inside the body.

A covariant definition of the center of inertia can be obtained if one keeps track of the (inertial) observer with respect to which it is defined. The covariant energy dipole moment defined by some observer with four-velocity $u^\mu$ relative to $\ucal S$ is given by
\begin{equation}\label{dipoledef}
D^\mu_u(\tau_u)=\int\ud_u^3x\,x^\mu T^{\alpha\beta}(x)u_\alpha u_\beta,
\end{equation}
where $\ud_u^3x\equiv\ud^4x\,\delta(x\cdot u-\tau_u)$ and $\tau_u$ are the Lorentz-invariant expressions for the volume measure and time coordinate defined by the observer. Since by construction $D^\mu_u(\tau_u)$ is a Lorentz four-vector, it can be used to define the coordinates of a physical point, referred to as centroid in the following
\begin{equation}\label{CMdefu}
R^\mu_u(\tau_u)\equiv\frac{D^\mu_u(\tau_u)}{P\cdot u},
\end{equation}
where $P\cdot u=\int\ud_u^3x\,T^{\alpha\beta}(x)u_\alpha u_\beta$ represents the Lorentz-invariant inertia defined by the observer. In particular, one has $R^\mu_u(\tau_u) u_\mu=\tau_u$ which indicates that the centroid worldline is parametrized by the observer's time coordinate. In a general Lorentz frame $\ucal S$, the centroid determined by Eq.~\eqref{CMdefu} and the center of inertia determined by Eq.~\eqref{CMdef} are usually different. They coincide however in the observer's frame $\ucal S_u$ where the four-velocity reduces to $u^\mu=(1,\uvec 0)$. This shows that the centroid represents in a covariant way the center of inertia defined by some observer. It is in this sense the Lorentz-invariant extension~\cite{Hoodbhoy:1998bt,Leader:2013jra} of the concept of center of inertia.

Similarly, the covariant generalized OAM tensor is defined as
\begin{equation}
M^{\alpha\beta}_{\text{orb},u}(\tau_u)\equiv\int\ud^3_ux\,M^{\mu\alpha\beta}_\text{orb}(x)u_\mu.
\end{equation}
By contraction with the four-velocity, we find that
\begin{equation}\label{Ru}
R^\mu_u(\tau_u)=\tau_u\,\frac{P^\mu}{P\cdot u}+\frac{M^{\mu\nu}_{\text{orb},u}(\tau_u)u_\nu}{P\cdot u}.
\end{equation}
The covariant internal OAM defined by the centroid reads
\begin{equation}\label{Su}
\ell^{\alpha\beta}_u(\tau_u)\equiv M^{\alpha\beta}_{\text{orb},u}(\tau_u)-R^\alpha_u(\tau_u)P^\beta+R^\beta_u(\tau_u)P^\alpha
\end{equation}
and provides an alternative definition of the centroid
\begin{equation}
\ell^{\alpha\beta}_u(\tau_u)u_\beta=0.
\end{equation}

\section{Symmetric energy-momentum tensor}\label{sec4}

In relativistic mechanics, one often assumes that the generalized total AM current has a pure orbital form~\cite{Misner:1974qy}
\begin{equation}
M^{\mu\alpha\beta}(x)\overset{!}{=} M^{\mu\alpha\beta}_\text{orb}(x)=x^\alpha T^{\mu\beta}(x)-x^\beta T^{\mu\alpha}(x).
\end{equation}
Two important consequences can then be derived from the conservation of this tensor:
\begin{enumerate}[label=\roman*)]
\item The EMT must be symmetric\footnote{It is also quite common to find the reverse argument in the literature. Assuming that the EMT is symmetric, one deduces that the corresponding orbital tensor must be conserved. The latter is then identified with the generalized total AM tensor.}
\begin{equation}
0=\partial_\mu M^{\mu\alpha\beta}(x)=T^{[\alpha\beta]}(x).
\end{equation}
The energy current coincides therefore with the four-momentum $\ucal J^\mu(x^0)=P^\mu$ and transforms as a time-independent Lorentz four-vector.
\item The generalized OAM is a time-independent Lorentz tensor 
\begin{equation}
M^{\alpha\beta}_\text{orb}(x^0)=M^{\alpha\beta}_{\text{orb},u}(\tau_u)=J^{\alpha\beta}
\end{equation}
identified with the generators of Lorentz transformations. 
\end{enumerate}
In the following, we discuss in some detail the implications for the center of inertia.

\subsection{Relativistic center-of-mass theorem}\label{sec41}

As one can see from Eq.~\eqref{CMR}, the conservation of the generalized OAM tensor forces the center of inertia to move along a straight line with constant velocity
\begin{equation}\label{inertialR}
R^\mu(x^0)=x^0\,\frac{P^\mu}{P^0}+Y^\mu\qquad \text{with}\qquad Y^\mu=\frac{J^{\mu 0}}{P^0},
\end{equation}
In particular, $Y^\mu=R^\mu(0)$ represents the position of the center of inertia at the initial time. While the generators of time translations $H=P^0$, spatial translations $\uvec P$ and rotations $\uvec J$ give the energy, linear and angular momentum of the system, the generators of  boosts $\uvec K$ give the initial energy dipole moment of the system
\begin{equation}
\uvec K=-\uvec Y P^0,
\end{equation}
and hence determine the initial position of the center of inertia. 

Differentiating Eq.~\eqref{inertialR} with respect to time shows that
\begin{equation}
P^\mu=P^0\dot R^\mu(x^0)
\end{equation}
which is nothing but the relativistic version of the CM theorem. Moreover, one obtains the relativistic version of K\"onig's first theorem
\begin{equation}
J^{\alpha\beta}=R^\alpha(x^0)\left[P^0\dot R^\beta(x^0)\right]-R^\beta(x^0)\left[P^0\dot R^\alpha(x^0)\right]+S^{\alpha\beta},
\end{equation}
when the generalized AM is decomposed into external and internal contributions.
\newline

Actually, the conservation of the generalized OAM tensor forces all the centroids to move along a straight line
\begin{equation}\label{inertialRu}
R^\mu_u(\tau_u)=\tau_u\,\frac{P^\mu}{P\cdot u}+Y^\mu_u\qquad \text{with}\qquad Y^\mu_u=\frac{J^{\mu \nu}u_\nu}{P\cdot u}.
\end{equation}
In order to determine the position of the centroid at some fixed time in $\ucal S$, we need to parametrize the corresponding worldline in terms of the time coordinate $x^0$ instead of the observer's time coordinate $\tau_u$. Since by definition $x^0=R^0_u(\tau_u)$, we find
\begin{equation}
x^0=\tau_u\,\frac{P^0}{P\cdot u}+Y^0_u
\end{equation}
which shows that the two time coordinates are related by a time dilation factor $\gamma_u=P^0/(P\cdot u)$ and a constant term $Y^0_u=J^{0\mu}u_\nu/(P\cdot u)=-\gamma_u(Y\cdot u)$ accounting for the relativity of simultaneity. Eliminating $\tau_u$ in Eq.~\eqref{inertialRu} in favor of $x^0$ leads to
\begin{equation}\label{Xu}
X^\mu_u(x^0)\equiv R^\mu_u(\tau_u(x^0))=\left(x^0-Y^0_u\right)\frac{P^\mu}{P^0}+Y^\mu_u.
\end{equation}
Since the centroid worldline is parametrized in terms of the time coordinate $x^0$, it is not surprising that manisfest Lorentz covariance is lost. This equation indicates two things: i) the dependence of $X^\mu_u$ on the four-velocity $u^\mu$ shows explicitly that the definition of a centroid is observer-dependent, and ii) all the simultaneous centroids move along parallel straight lines with the same constant velocity.

Because of the special role played by the CM frame $\ucal S_\star$, let us consider the corresponding centroid. Using $u^\mu_\star=P^\mu /M$ in Eq.~\eqref{inertialRu}, we find that the position in $\ucal S$ of the proper centroid is given by
\begin{equation}
R^\mu_\star(\tau)=\tau\,\frac{P^\mu}{M}+Y^\mu_\star\qquad\text{with}\qquad Y^\mu_\star=\frac{J^{\mu\nu}P_\nu}{M^2}.
\end{equation}
This expression coincides with Eq.~\eqref{CMpos} since $Y^\mu_\star=-K^\mu_\star/M$, showing that the CM is nothing but the proper centroid.

\subsection{Supplementary spin condition}\label{sec42}

The inertial motion of the centroids~\eqref{Xu} implies that the corresponding external and internal parts of the generalized AM tensor are separately time-independent
\begin{equation}\label{Lcons}
L^{\alpha\beta}_u=X^\alpha_u(x^0)P^\beta-X^\beta_u(x^0)P^\alpha=Y^\alpha_u P^\beta-Y^\beta_u P^\alpha,\qquad S^{\alpha\beta}_u=J^{\alpha\beta}-L^{\alpha\beta}_u.
\end{equation}
When the generalized AM tensor is purely orbital, the internal part reads $S^{\alpha\beta}_u=\ell^{\alpha\beta}_u$. Choosing the centroid as the reference point leads then to the equation
\begin{equation}\label{SSC}
S^{\alpha\beta}_uu_\beta=0,
\end{equation}
which is known in the literature of classical spinning bodies under the name of supplementary spin condition (SSC), see e.g.~\cite{Costa:2014nta} and references therein. In particular, choosing the CM as the reference point corresponds to imposing the Tulczyjew-Dixon SSC $S^{\alpha\beta}_\star P_\beta=0$~\cite{Tulczyjew:1959,Dixon:1964,Costa:2014nta}. 

One can think of a SSC as the requirement that covariant boost generators do not contribute to the internal part of the generalized AM. Accordingly, we can write
\begin{equation}
S^{\alpha\beta}_u=-\epsilon^{\alpha\beta\mu\nu}S_{u\mu}u_\nu
\end{equation}
owing to Eq.~\eqref{Jdec}. In other words, the tensor $S^{\alpha\beta}_u$ represents the internal AM of the system defined by an observer with four-velocity $u^\mu$ relative to $\ucal S$. Working out explicitly the expression for $S^{\alpha\beta}_u$ in Eq.~\eqref{Lcons} with the centroid given by Eq.~\eqref{inertialRu}, we find that it is related to the Pauli-Luba\'nski pseudo-vector as follows
\begin{equation}\label{Su}
S^{\alpha\beta}_u=-\frac{\epsilon^{\alpha\beta\mu\nu}W_\mu u_\nu}{P\cdot u}.
\end{equation}
The internal AM defined by an observer with four-velocity $u^\mu$ is then given by
\begin{equation}\label{intAM}
S^\mu_u=(\delta^\mu_\nu-u^\mu u_\nu)\,\frac{W^\nu}{P\cdot u}.
\end{equation}
For $u^\mu_\star=P^\mu/M$, we naturally recover $J^\mu_\star=S^\mu_\star=W^\mu/M$.

\subsection{Transverse shifts}\label{sec43}

Using Eqs.~\eqref{Qshift} and~\eqref{refpoint}, one can see that the two conserved quantities $L^{\mu 0}_u/P^0$ and $S^{\mu 0}_u/P^0$ respectively represent the initial position of the centroid and the shift of the center of inertia relative to it 
\begin{align}
X^\mu_u(0)=\frac{L^{\mu 0}_u}{P^0},\qquad\qquad
Q^\mu_u\equiv R^\mu(x^0)-X^\mu_u(x^0)=\frac{S^{\mu 0}_u}{P^0}.\label{covshift}
\end{align}
Combined with Eq.~\eqref{Su} for the internal AM, we get
\begin{equation}
Q^\mu_u=-\frac{\epsilon^{\mu 0\alpha\beta}W_\alpha u_\beta}{P^0 (P\cdot u)}
\end{equation}
which shows that the shift in $\ucal S$ is spatial and transverse to both $\uvec W$ and $\uvec u$.

More generally, one can consider the shift\footnote{Recently, the importance of these transverse shifts or ``side jumps'' for the Lorentz invariance has been stressed in the context of the chiral kinetic theory~\cite{Chen:2014cla,Chen:2015gta}.} between any two centroids at any fixed time $x^0$
\begin{equation}
Q^\mu_{uv}\equiv X^\mu_u(x^0)-X^\mu_v(x^0)=\frac{S^{\mu 0}_v-S^{\mu 0}_u}{P^0}=Q^\mu_v-Q^\mu_u.
\end{equation}
This shift can also be written as
\begin{equation}
Q^\mu_{uv}=-\frac{\epsilon^{\mu 0\alpha\beta}W_\alpha V_\beta}{(P^0)^2},
\end{equation}
which shows that $Q^\mu_{uv}$ is spacelike and orthogonal to both the Pauli-Luba\'nski pseudo-vector $W^\mu$ and the relative velocity four-vector $V^\mu\equiv \gamma_v v^\mu-\gamma_u u^\mu$.

\subsection{M\o ller's disk}\label{sec44}

Instead of looking at the position of the centroids at some fixed time in $\ucal S$, we can look at their position at some fixed proper time $\tau$. A contraction of Eq.~\eqref{inertialRu} with $u^\mu_\star$ allows us to relate the parameter $\tau_u$ with the proper time $\tau$
\begin{equation}
\tau=\tau_u\,\frac{M}{P\cdot u}+\frac{Y_u\cdot P}{M},
\end{equation}
leading to
\begin{equation}
\widetilde X^\mu_u(\tau)\equiv R^\mu_u(\tau_u(\tau))=\left(\tau-\frac{Y_u\cdot P}{M}\right)\frac{P^\mu}{M}+Y^\mu_u.
\end{equation}
This clearly shows that all the centroids have the same four-velocity $u^\mu_\star$.

For the proper shift of a centroid relative to the CM $\widetilde Q^\mu_u\equiv \widetilde X^\mu_u(\tau)-R^\mu_\star(\tau)$, we find that it is given by
\begin{equation}\label{covshift}
\widetilde Q^\mu_u=-\frac{S^{\mu\nu}_u P_\nu}{M^2}=\frac{S^{\mu\nu}_\star u_\nu}{P\cdot u}=-\frac{\epsilon^{\mu\nu\alpha\beta}u_\nu S_{\star\alpha} P_\beta}{M (P\cdot u)}.
\end{equation}
In other words, the CM is a center of inertia in all Lorentz frames only when the spin of the system vanishes $J^\mu_\star=S^\mu_\star=0$. When the latter is nonzero, the proper shift is orthogonal to $P^\mu$, $u^\mu$ and $S^\mu_\star$, and therefore is the same in both the observer's frame $\ucal S_u$ and the CM frame $\ucal S_\star$. In the CM frame, our expression~\eqref{covshift} reduces to~\cite{Moller:1949,Moller:1949bis}
\begin{equation}
\widetilde{\uvec Q}_u\overset{\ucal S_\star}{=}-\frac{\uvec v\times\uvec S_0}{M},
\end{equation}
where $\uvec v=\uvec u/u^0$ is the velocity of the observer relative to $\ucal S_\star$, and the spin is given by $S^\mu_\star\overset{\ucal S_\star}{=}(0,\uvec S_0)$. By varying the relative velocity $\uvec v$, we obtain the position of all the centroids relative to the CM. They form the so-called M\o ller disk~\cite{Moller:1949,Moller:1949bis}, which is orthogonal to the spin of the system $\uvec S_0$, has radius
\begin{equation}
R_\text{M\o ller}=\frac{|\uvec S_0|}{M},
\end{equation}
and whose geometric center coincides with the CM. The disk is at rest in $\ucal S_\star$ and moves as a rigid body with constant velocity in any other frame $\ucal S$. Note that M\o ller's disk is an open set since $|\uvec v|<1$ for massive observers.
\newline

M\o ller's disk plays an important role for extended bodies because it provides a lower bound on the dimensions of a classical system. If one assumes that i) the energy density in the convex spherical hull of the system is positive in any frame and ii) the AM is purely orbital, then M\o ller's disk lies fully inside the convex spherical hull~\cite{Moller:1949,Moller:1949bis,Costa:2014nta}. In other words, a classical system with spin $|\uvec S_0|$ and mass $M$ necessarily has a typical radial dimension $r$ larger than M\o ller's radius
\begin{equation}\label{Mollerbound}
r> \frac{|\uvec S_0|}{M}.
\end{equation}
This is a purely relativistic effect. Indeed, reinstating the factor of $1/c$ in the above expression shows that the lower bound vanishes in the non-relativistic limit $c\to\infty$. The appearance of M\o ller's radius in a relativistic theory can also be understood somewhat intuitively. If we imagine that the mass $M$ of an extended body can be concentrated at one of its physical points $P$, the spin of the system will then be given by the rotation of $P$ about the axis parallel to $\uvec S_0$ passing through the CM. Denoting the distance between $P$ and the rotation axis by $r$ and the angular velocity by $\omega$, one finds that $|\uvec S_0|=M\omega r^2$. Relativity imposes that the velocity of $P$ cannot exceed the speed of light $\omega r\leq c$. It then follows that $r$ must be larger than M\o ller's radius.


\section{Asymmetric energy-momentum tensor}\label{sec5}

In the literature, it is often claimed that the fundamental EMT must be symmetric. This claim is essentially based on two arguments: i) in General Relativity the gravitational EMT of matter, defined as the source of the gravitational field, appears to be symmetric, and ii) conservation of the orbital tensor forces the EMT to be symmetric~\cite{Misner:1974qy}. These arguments are however by no means actual proofs~\cite{Lorce:2015lna}. 

General Relativity does not require the fundamental EMT to be symmetric, but simply indicates that only a symmetric part of it couples to gravitation. Note also that the symmetry of this gravitational part follows from the assumption that the torsion of spacetime vanishes. When the latter condition is relaxed, the gravitational EMT turns out in general to be asymmetric~\cite{Hehl:1976vr}. Regarding the second argument, the assumption that the orbital tensor is conserved comes from Classical Mechanics, where there exists only one form of AM. As soon as one introduces the intrinsic AM of Quantum Mechanics, which is a new form of AM, there is no reason to maintain this assumption. In conclusion, we do not see any fundamental reason for requiring the EMT to be symmetric.
\newline

In field theory, the generic Lorentz transformation law of a multicomponent relativistic field reads in the active viewpoint
\begin{equation}
\phi(x)\mapsto \phi'(x)=M[\Lambda]\,\phi(\Lambda^{-1}x),
\end{equation}
where $M[\Lambda]$ is a matrix acting on the field components. It then follows from the canonical formalism that the associated conserved current naturally receives two contributions
\begin{equation}\label{Mtot}
M^{\mu\alpha\beta}(x)=M^{\mu\alpha\beta}_\text{orb}(x)+ M^{\mu\alpha\beta}_\text{int}(x).
\end{equation}
Beside the orbital (or extrinsic) tensor associated with the transformation of the point, there is an intrinsic tensor $M^{\mu\alpha\beta}_\text{int}=-M^{\mu\beta\alpha}_\text{int}$ associated with the mixing of the field components.

From the conservation of $M^{\mu\alpha\beta}(x)$, one concludes that the asymmetry of the EMT is related to the non-conservation of the intrinsic tensor~\cite{Papetrou:1949,Truesdell:1960,Eringen:1976,Hehl:1976vr}
\begin{equation}\label{local}
T^{[\alpha\beta]}(x)=\partial_\mu M^{\mu\alpha\beta}_\text{orb}(x)=-\partial_\mu M^{\mu\alpha\beta}_\text{int}(x).
\end{equation}
Focusing on the spatial currents $\alpha\beta=ij$, this means that orbital and intrinsic AM are not separately conserved in general. They can be converted into each other owing to spin-orbit couplings, leading to an asymmetric EMT. This phenomenon is illustrated by e.g. the Einstein-de Haas effect~\cite{Einstein:1915,Frenkel:1979} routinely used to measure the gyromagnetic ratio of atoms and molecules~\cite{Barnett:1953}. 

Because of the non-conservation of the orbital tensor $\partial_\mu M^{\mu\alpha\beta}_\text{orb}(x)\neq 0$, the results of Section~\ref{sec4} should not be expected to hold in general anymore. In particular, the centroids are not forced to move along straight lines with constant velocity, so that there can be an exchange of generalized AM between the external and internal parts. This is annoying since it jeopardizes the usefulness of the concept of center of inertia in field theory, and hence the clear connection with Classical Mechanics.

\subsection{Belinfante-Rosenfeld procedure}

There exists a freedom in the definition of the EMT and the generalized AM tensor. Starting from some couple $T^{\mu\nu}(x),M^{\mu\alpha\beta}(x)$ a whole family of alternative couples can be defined by
\begin{align}
T^{\mu\nu}_G(x)&=T^{\mu\nu}(x)+\partial_\lambda G^{\lambda\mu\nu}(x),\label{TBel}\\
M_G^{\mu\alpha\beta}(x)&=M^{\mu\alpha\beta}(x)+\partial_\lambda[x^\alpha G^{\lambda\mu\beta}(x)-x^\beta G^{\lambda\mu\alpha}(x)],
\end{align}
where the superpotential is required to satisfy a symmetry property $G^{[\lambda\mu]\nu}(x)=0$, so that the new tensors remain conserved and lead to the same Poincar\'e generators as the original ones, provided that surface terms vanish at spatial infinity
\begin{align}
P^\mu&=\int\ud^3_ux\,T^{\lambda\mu}_G(x)u_\lambda=\int\ud^3_ux\,T^{\lambda\mu}(x)u_\lambda,\\
J^{\alpha\beta}&=\int\ud^3_ux\,M^{\mu\alpha\beta}_G(x)u_\mu=\int\ud^3_ux\,M^{\mu\alpha\beta}(x)u_\mu.
\end{align}
The effect of the superpotential consists therefore in a mere relocalization of the linear and angular momentum distributions~\cite{Hehl:1976vr}. In particular, we observe that the superpotential operates a transfer between orbital and intrinsic tensors
\begin{align}
M_{G,\text{orb}}^{\mu\alpha\beta}(x)&=M_\text{orb}^{\mu\alpha\beta}(x)+[x^\alpha \partial_\lambda G^{\lambda\mu\beta}(x)-x^\beta \partial_\lambda G^{\lambda\mu\alpha}(x)],\\
M_{G,\text{int}}^{\mu\alpha\beta}(x)&= M_\text{int}^{\mu\alpha\beta}(x)- G^{\mu[\alpha\beta]}(x),
\end{align}
and therefore shifts the position of the centroids by
\begin{equation}\label{CMshift}
Q^\mu_{G,u}(\tau_u)\equiv R^\mu_{G,u}(\tau_u)-R^\mu_u(\tau_u)=-\frac{1}{P\cdot u}\int\ud^3_ux\, G^{\mu \alpha\beta}(x)u_\alpha u_\beta.
\end{equation}

Belinfante and Rosenfeld~\cite{Belinfante:1939,Belinfante:1940,Rosenfeld:1940} noticed that the new intrinsic part can be set to zero using the particular choice
\begin{equation}\label{GBel}
G^{\lambda\mu\nu}_\text{Bel}(x)=\frac{1}{2}\left[ M^{\lambda\mu\nu}_\text{int}(x)+M^{\mu\nu\lambda}_\text{int}(x)+ M^{\nu\mu\lambda}_\text{int}(x)\right].
\end{equation}
Since the Belinfante-Rosenfeld generalized AM tensor is by construction purely orbital
\begin{equation}
M^{\mu\alpha\beta}_\text{Bel}(x)=x^\alpha T^{\mu\beta}_\text{Bel}(x)-x^\beta T^{\mu\alpha}_\text{Bel}(x),
\end{equation}
we fall back to the case studied in Section~\ref{sec4}. In particular, it follows from the conservation of the Belinfante-Rosenfeld generalized AM tensor $\partial_\mu M^{\mu\alpha\beta}_\text{Bel}(x)=0$ that the Belinfante-Rosenfeld EMT is symmetric $T^{[\alpha\beta]}_\text{Bel}(x)=0$, and the Belinfante-Rosenfeld centroids move along straight lines with constant velocity as given by Eq.~\eqref{inertialRu}
\begin{equation}
R^\mu_{\text{Bel},u}(\tau_u)=\tau_u\,\frac{P^\mu}{P\cdot u}+Y^\mu_u\qquad \text{with}\qquad Y^\mu_u=\frac{J^{\mu \nu}u_\nu}{P\cdot u}.
\end{equation}

Because of its symmetry, the Belinfante-Rosenfeld EMT is often considered in the literature to be more fundamental than the canonical one. We disagree with this point of view for several reasons. The first reason is that, as argued above, there does not exist any fundamental reason which requires the EMT to be symmetric. Second, the relocalization of linear and angular momentum distributions in the Belinfante-Rosenfeld procedure is totally ad hoc and does not follow from a canonical approach. The last reason is also the most important one. The Belinfante-Rosenfeld procedure is compatible with General Relativity, because the assumed covariance under diffeomorphisms implies that only the charges (i.e. integrated current densities) can be considered as physical. In other words, the relocalization of the linear and angular momentum distributions is considered to be unphysical and hence harmless. Orbital and intrinsic forms of AM seem therefore indistinguishable in the context of General Relativity. This is at odds with Quantum Mechanics, where orbital and intrinsic forms of AM are fundamentally different and distinguishable, as confirmed by numerous experiments. Linear and angular momentum distributions cannot be relocalized at will. In our view, the Belinfante-Rosenfeld procedure corresponds actually to an effective description of the system, where the intrinsic contribution is mimicked by a modification of the distribution of energy and momentum. For further discussions about canonical and Belinfante-Rosenfeld tensors, see~\cite{Leader:2013jra}.

\subsection{Intrinsic energy dipole moment}

It follows from Eq.~\eqref{Mtot} that the generalized AM tensor can be covariantly decomposed as
\begin{equation}
J^{\alpha\beta}=M^{\alpha\beta}_{\text{orb},u}(\tau_u)+M^{\alpha\beta}_{\text{int},u}(\tau_u),
\end{equation}
where the separate orbital and intrinsic contributions are in general time-dependent. The Belinfante-Rosenfeld centroids are therefore shifted from the original ones
\begin{equation}\label{intshift}
R^\mu_{\text{Bel},u}(\tau_u)=R^\mu_u(\tau_u)+Q^\mu_{\text{int},u}(\tau_u)
\end{equation}
by a term depending on the intrinsic part\footnote{For this reason, the Belinfante-Rosenfeld CM $R^\mu_{\text{Bel},\star}$ has been coined ``center of mass and spin'' by Medina and Stephany~\cite{Medina:2014gra,Medina:2014bwa,Medina:2017mcd}. 
}
\begin{equation}
Q^\mu_{\text{int},u}(\tau_u)=\frac{M^{\mu\nu}_{\text{int},u}(\tau_u)u_\nu}{P\cdot u}.
\end{equation}
Accordingly, the quantity
\begin{equation}\label{intdip}
D^\mu_{\text{int},u}(\tau_u)\equiv M^{\mu\nu}_{\text{int},u}(\tau_u)u_\nu=-K^\mu_{\text{int},u}(\tau_u)
\end{equation}
can be interpreted as an intrinsic covariant energy dipole moment. Like spin, it is an intrinsic property of elementary particles. Its time dependence indicates that, just like AM, some energy dipole moment can be exchanged between orbital and intrinsic parts. It is because of this exchange that the relativistic CM theorem does not hold in general. Since the covariant centroid velocity is given by
\begin{equation}\label{covvelocity}
\dot R^\mu_u(\tau_u)=\frac{\ucal J^\mu_u(\tau_u)}{P\cdot u},
\end{equation}
this amounts to say that the covariant energy current $\ucal J^\mu_u(\tau_u)\equiv\int\ud^3_ux\,T^{\mu\nu}(x)u_\nu$ is in general time-dependent.

\subsection{Intrinsic spin conjecture}

Instead of following the Belinfante-Rosenfeld procedure and demanding that the new intrinsic tensor vanishes $M^{\mu\alpha\beta}_{G,\text{int}}(x)\overset{!}{=}0$, we observe from Eq.~\eqref{intshift} that we can impose a weaker condition to recover the validity of the relativistic CM theorem. We simply require that all the intrinsic energy dipole moments must vanish
\begin{equation}
D^\mu_{\text{int},u}(\tau_u)=\int\ud^3_ux\,M^{\lambda\mu\nu}_\text{int}(x)u_\lambda u_\nu\overset{!}{=}0,\qquad \forall\, u^\mu\text{ timelike}.
\end{equation} 
Combined with the antisymmetry in the last two indices $M^{\mu[\alpha\beta]}_\text{int}=0$, this amounts to requiring the intrinsic tensor to be totally antisymmetric, and hence expressible in terms of a sole pseudo-vector $\ucal A_\lambda(x)\equiv\frac{1}{3!}\,\epsilon_{\lambda\mu\alpha\beta}\,M^{\mu\alpha\beta}_\text{int}(x)$ as follows
\begin{equation}\label{ISC}
M^{\mu\alpha\beta}_\text{int}(x)\overset{!}{=}\epsilon^{\mu\alpha\beta\lambda}\,\ucal A_\lambda(x).
\end{equation}
For example, the pseudo-vector is related to the axial-vector bilinear $\ucal A^\mu_{1/2}(x)=\frac{1}{2}\,\overline\psi(x)\gamma^\mu\gamma_5\psi(x)$ in Dirac theory, and to the Chern-Simons current $\ucal A^\mu_1(x)=-\frac{1}{3}\,\epsilon^{\mu\nu\alpha\beta}A_\nu(x)F_{\alpha\beta}(x)$ in Maxwell theory\footnote{For a gauge-invariant formulation, see~\cite{Guo:2012wv}.}. We will refer to the condition in Eq.~\eqref{ISC} as the intrinsic spin conjecture (ISC), since it enforces the intrinsic part to contribute to the rotation generators only
\begin{equation}
K^\mu_{\text{int},u}(\tau_u)=0,\qquad J^\mu_{\text{int},u}(\tau_u)=\left(\delta^\mu_\nu-u^\mu u_\nu\right)s^\nu_u(\tau_u)
\end{equation}
with $s^\mu_u(\tau_u)=\int\ud^3_ux\,\ucal A^\mu(x)$. In other words, elementary particles are characterized by some intrinsic AM but no intrinsic energy dipole moment\footnote{For a discussion about particles with non-vanishing intrinsic energy dipole moment, see~\cite{Ivezic:2013nva,Salesi:2015ita}.}~\cite{Anandan:1999ig}.

Differentiating Eq.~\eqref{local} and using the antisymmetry of the intrinsic tensor in its first two indices implies that the EMT satisfies another conservation law
\begin{equation}\label{spinconj}
\partial_\nu T^{\mu\nu}(x)=0
\end{equation}
beside the usual one $\partial_\mu T^{\mu\nu}(x)=0$. Clearly, this condition is fulfilled when the EMT is symmetric, but it is not a necessary requirement. It follows from this new conservation law that the covariant energy current is time-independent $\dot{\ucal J}^\mu_u(\tau_u)=0$, confirming that the relativistic CM theorem holds, see Eq.~\eqref{covvelocity}. 

The validity of the CM theorem is sufficient to recover the results obtained in Sections~\ref{sec42}-\ref{sec44}. The only difference is that now the internal part can be further decomposed as follows
\begin{equation}
S^{\alpha\beta}_X(\tau_u)=\ell^{\alpha\beta}_{Xu}(\tau_u)+s^{\alpha\beta}_u(\tau_u),
\end{equation}
where the orbital and intrinsic contributions are given by
\begin{align}
\ell^{\alpha\beta}_{Xu}(\tau_u)&=\int\ud^3_ux\left[(x^\alpha-X^\alpha(\tau_u))\, T^{\mu\beta}(x)-(x^\beta-X^\beta(\tau_u))\, T^{\mu\alpha}(x)\right]u_\mu,\\
s^{\alpha\beta}_u(\tau_u)&=\int\ud^3_ux\,M^{\mu\alpha\beta}_\text{int}(x)u_\mu=-\epsilon^{\alpha\beta\mu\nu}s_{u\mu}(\tau_u)u_\nu.
\end{align}
For later convenience, we left open the possibility to use a reference point $X^\mu(\tau_u)$ different from the centroid $R^\mu_u(\tau_u)$. Because of the ISC we have the relation $s^{\alpha\beta}_u(\tau_u)u_\beta=0$, and so the shift of the centroid relative to the reference point is simply given by
\begin{equation}
R^\mu_u(\tau_u)-X^\mu(\tau_u)=\frac{S^{\mu\nu}_X(\tau_u)u_\nu}{P\cdot u}=\frac{\ell^{\mu\nu}_{Xu}(\tau_u)u_\nu}{P\cdot u}.
\end{equation}
Moreover, if $X^\mu_v(\tau_u)$ is the centroid defined by some observer with four-velocity $v^\mu$ relative to the Lorentz frame $\ucal S$, we can write
\begin{equation}
\ell^{\alpha\beta}_{vu}(\tau_u) v_\beta=\epsilon^{\alpha\beta\mu\nu}v_\beta s_{u\mu}(\tau_u)u_\nu,
\end{equation}
since the SSC $S^{\alpha\beta}_vv_\beta=0$~\eqref{SSC} is satisfied by definition of the centroid.

\subsection{Elementary particles}\label{elementaryparticle}

A free elementary particle has by definition no internal structure and is usually pictured as a pointlike object. Where is this point located? It is often thought that the answer is either the center of inertia or the reference point, but the correct one is the CM, because it is the only physical point (i.e. with components transforming as a Lorentz four-vector) unambiguously defined by the system. In the CM frame, there is no ambiguity since the particle is at rest and the CM coincides with the center of inertia. Moreover, the AM is purely internal and hence independent of the reference point. In this frame, we therefore expect that the EMT takes the form $T^{\mu\nu}(x)\overset{\ucal S_\star}{=}\eta^{0\mu}\eta^{0\nu}\,\delta^{(4)}(x-R_\star(\tau))$, where $\eta_{\mu\nu}=\text{diag}(+1,-1,-1,-1)$ is the Minkowski metric, leading to $\ell^{\alpha\beta}_\star(x^0)\overset{\ucal S_\star}{=}0$ and hence $J^\mu_\star\overset{\ucal S_\star}{=}s^\mu_\star(x^0)$. 

The situation is more complicated for a moving particle. Indeed, the canonical description of a moving state is obtained from a rotationless boost applied to a state at rest~\cite{Polyzou:2012ut}. Since the definition of rotationless boosts depends on the observer, one may expect that the reference point naturally associated with this moving state (hereafter called canonical reference point) will be observer-dependent\footnote{Thomas precession can then be understood as a pure kinematic effect arising from a continuous change of canonical reference point~\cite{Moller:1949,Muller:1992,Rebilas:2015}.}. The naive guess would be that the canonical reference point $\ucal R^\mu(x^0)$ does coincide with the center of inertia $R^\mu(x^0)$. A careful calculation shows that it lies in fact somewhere in between the center of inertia and the CM, see Appendix~\ref{AppB}. As a result, the internal AM will receive in general both orbital and intrinsic contributions $S^\mu_{\ucal R}=\ell^\mu_{\ucal R}(x^0)+s^\mu(x^0)$, because the canonical reference point appears to be shifted sideways relative to the CM when polarization is not aligned with momentum. This simple phenomenon explains why the longitudinal spin decomposition is frame-independent whereas the transverse spin decomposition is not. It can be illustrated with the example of a Dirac particle.

A free elementary spin-$\tfrac{1}{2}$ particle with mass $m$, momentum $\uvec p$, energy $E=\sqrt{\uvec p^2+m^2}$, and canonical polarization $s=\uparrow,\downarrow$ can be described by the positive-energy Dirac spinor
\begin{equation}
\psi_{\uvec p,s}(x)=u(\uvec p,s)\,e^{-ip\cdot x}.
\end{equation}
In the standard representation with spin quantization axis along the $z$-direction, the momentum-space spinor reads
\begin{equation}
u(\uvec p,s)=\sqrt{E+m}\begin{pmatrix}\chi_s\\\frac{\uvec p\cdot\uvec \sigma}{E+m}\,\chi_s\end{pmatrix}\qquad \text{with}\qquad \chi_\uparrow=\begin{pmatrix}1\\0\end{pmatrix},\, \chi_\downarrow=\begin{pmatrix}0\\1\end{pmatrix}.
\end{equation}
It contains\footnote{In the language of first quantization, the upper and lower components of the Dirac spinor are simultaneous eigenstates of the longitudinal intrinsic AM operator $\hat S_z=\frac{1}{2}\,\gamma^3\gamma_5$ and orbital AM operator $\hat L_z=\gamma^0\left(p_R\,\frac{\partial}{\partial p_R}-p_L\,\frac{\partial}{\partial p_L}\right)$. For a state with polarization $s=\uparrow$ corresponding to the eigenvalue of total AM $j_z=+\frac{1}{2}$, the upper components have eigenvalues $s_z=+\frac{1}{2}$ and $l_z=0$, and the lower components lead to a superposition of $s_z=+\frac{1}{2}$ and $l_z=0$ with $s_z=-\frac{1}{2}$ and $l_z=+1$.} both an $s$-wave (upper two components) and a $p$-wave (lower two components). The AM sum rule reads
\begin{equation}
\frac{1}{2}=\frac{1}{2E}\,\overline u(\uvec p,\uparrow)\left[\frac{1}{2}\,\gamma^3\gamma_5+\gamma^0\left(p_R\,\frac{\partial}{\partial p_R}-p_L\,\frac{\partial}{\partial p_L}\right)\right]u(\uvec p,\uparrow),
\end{equation}
where $p_{R,L}=p^1\pm ip^2$. The first term represents the intrinsic contribution and the second term represents the OAM contribution defined relative to the canonical reference point, which is set at the origin of the coordinate system so that the external AM vanishes. In the rest frame of the particle $\uvec p=\uvec 0$, the $p$-wave vanishes and the AM is purely intrinsic. When the particle is moving $\uvec p\neq\uvec 0$, the $p$-wave sets in and provides the orbital contribution to the AM. As the momentum increases, the contribution from the $p$-wave becomes more and more important, and ultimately reaches the same weight as the $s$-wave in the ultrarelativistic limit $|\uvec p|\to\infty$. As long as the momentum remains parallel to spin, the AM will be purely intrinsic because the orbital contribution of the $p$-wave vanishes (this corresponds to $l=1$ but $l_z=0$). As soon as the momentum develops a component orthogonal to spin, the intrinsic contribution decreases and is compensated by an increase of the orbital contribution. In the limit of infinitely large transverse momentum, the intrinsic contribution vanishes (or better averages out between $s$ and $p$-waves) so that the AM becomes purely orbital. In this ultrarelativistic limit, the value $\tfrac{1}{2}$ for the AM arises from the average of an $s$-wave and a $p$-wave with equal weight. This generalizes straightforwardly to spin-$j$ particles with maximal polarization orthogonal to momentum, where the AM in the ultrarelativistic regime arises from the average of $2j$ partial waves with equal weight $\frac{1}{2j_\text{max}+1}\sum_{l_\text{max}=0}^{2j_\text{max}}l_\text{max}=j_\text{max}$, with $l_\text{max}$ and $j_\text{max}$ denoting the maximal projections of orbital and total AM along the quantization axis, respectively.
\newline

Why does the canonical reference point depend in general on the observer? The reason is simply because we are considering eigenstates with a fixed (quantized) eigenvalue of the \emph{internal} AM.

We have seen that $J^{\alpha\beta}$ behaves as an antisymmetric rank-2 Lorentz tensor. It then follows that, like for the components of the electromagnetic field, the value of $J^z$ will depend on the frame, see Eq.~\eqref{JLT}. There is no way to preserve in general the value of $J^z$ under a canonical boost. Unlike $J^{\alpha\beta}$, the external and internal parts need not transform individually as antisymmetric rank-2 Lorentz tensors. They will behave as Lorentz tensors only if the coordinates of the reference point $X^\mu$ transform as a Lorentz four-vector, i.e. only when the reference point is a physical point that does not depend on the observer. If we choose a reference point whose components do not transform as a Lorentz four-vector, i.e. whose definition depends on observer like e.g. the center of inertia, there is a possibility to preserve the value of the internal AM. 

The canonical reference point $\ucal R^\mu$, derived in Appendix B, is precisely the one for which the internal AM $S^z_{\ucal R}$ stays constant under a canonical Lorentz transformation from the rest frame. In the rest frame, the canonical reference point coincides with the CM. When the momentum of the system is increased by a canonical boost, the canonical reference point shifts away from the CM and generate an additional OAM contribution which ensures that the internal AM $S^z_{\ucal R}$ remains the same in any frame. In other words, the appearance of a $p$-wave in the Dirac plane-wave solutions can be understood as a relativistic quantum-mechanical effect.

Usually, one identifies for convenience the origin $O$ of the coordinate system with the canonical reference point $\ucal R$ to get rid of the external AM contribution represented by the first term on the right-hand side of Eq.~\eqref{genexpr}. This amounts to identifying the total AM with the internal part $J^z=S^z_O=S^z_{\ucal R}$. In the above discussion, it was essential to distinguish total AM from the internal part, since otherwise one would have concluded that the origin gets shifted under Lorentz boosts!

\section{Recap of the generalized angular momentum decomposition}

As we have seen, the generalized AM can be decomposed in several ways. In order to clarify the global picture and the terminology, we pause for a moment and summarize the general structure of this decomposition.
\begin{enumerate}[label=\arabic*)]
\item By choosing an observer, the generalized AM can be decomposed into \emph{boost} and \emph{angular momentum} contributions.
\item By choosing a reference point (or pivot), boosts and AM can further be decomposed into \emph{external} (or translation-dependent) and \emph{internal} (or translation-independent) contributions.
\item Boosts are purely \emph{orbital} (or extrinsic), whereas AM receives both \emph{orbital} and \emph{intrinsic} contributions.
\item When the system consists of several types of constituents, all the above contributions can further be decomposed according to the constituent types.
\end{enumerate}
We also summarize the terminology
\begin{enumerate}[label=\arabic*)]
\item The \emph{spin} of a system is the AM defined by an observer sitting in the CM frame. It is purely internal and decomposes into orbital and intrinsic contributions when seen from a generic Lorentz frame.
\item The \emph{center of mass} is the particular reference point for which boosts defined by an observer sitting in the CM frame become purely external. It coincides with the center of inertia (or energy) in the CM frame.
\item The \emph{canonical reference point} is the pivot about which the internal AM remains constant under canonical boosts.
\end{enumerate}

\section{Light-front formalism}\label{sec6}

Dirac showed that there exist several forms in which relativistic dynamics can be formulated~\cite{Dirac:1949cp}. So far, we have considered the so-called instant form, which is naturally associated with massive observers. In the context of high-energy physics, where mass effects can often be neglected, it is the so-called front form that appears to be more useful~\cite{Brodsky:1997de}.

The covariant formulation used in this paper allows us to easily transpose the results obtained in instant form to the light-front (LF) formalism. In the covariant formulation of instant form, an observer is characterized by a timelike unit four-vector $u^\mu$ representing its four-velocity relative to the Lorentz frame $\ucal S$. In the covariant formulation of front form we obtain the same expressions, but this time an observer will be characterized by a lightlike four-vector $n^\mu$~\cite{Carbonell:1998rj}. In some sense, using the front form of dynamics amounts to adopting the point of view of a massless observer, formalizing therefore Einstein's thought experiment of riding a photon. Note that instant and front forms coincide in the infinite-momentum frame since
\begin{equation}
\lim_{|\uvec u|\to \infty}\frac{u^\mu}{u^0}\propto n^\mu\qquad \text{with}\qquad u^0=\sqrt{\uvec u^2+1}.
\end{equation}

\subsection{Light-front components}\label{LFcomponent}

For later convenience, we introduce the dual lightlike four-vector $\bar n^\mu$ satisfying by definition $\bar n\cdot n=1$. The two lightlike four-vectors can then be used to perform the Sudakov decomposition of any four-vector
\begin{equation}
a^\mu=a^+\bar n^\mu+a^-n^\mu+a^\mu_\perp,
\end{equation}
where the LF components are defined as $a^+\equiv a\cdot n$ and $a^-\equiv a\cdot \bar n$, and the transverse components by $a^\mu_\perp\equiv\eta^{\mu\nu}_\perp a_\nu$ with the transverse projector $\eta^{\mu\nu}_\perp=\eta^{\mu\nu}-n^\mu \bar n^\nu-\bar n^\mu n^\nu$. 

The scalar product of two four-vectors can be written as
\begin{equation}
a\cdot b=a^+ b^-+a^-b^++a_\perp\cdot b_\perp.
\end{equation}
In the particular case of four-position and four-momentum, we find $x\cdot P=x^+ P^-+x^-P^++x_\perp\cdot P_\perp$. If we choose $x^+$ to the represent the LF time coordinate, then the LF energy component is represented by $P^-$, whereas the LF (longitudinal) components of position and momentum are represented by $x^-$ and $P^+$, respectively. Note that for a massive system, we have $P^\pm>0$.

In the LF formalism, the generalized AM tensor can be written as
\begin{equation}\label{JdecLF}
J^{\alpha\beta}=-K^\alpha_n \bar n^\beta+K^\beta_n \bar n^\alpha-\epsilon^{\alpha\beta\mu\nu}J_{n\mu} n_\nu,
\end{equation}
where the covariant LF boost and rotation generators are defined by
\begin{equation}
K^\mu_n\equiv -J^{\mu\nu}n_\nu\qquad\text{and}\qquad J^\mu_n\equiv\tfrac{1}{2}\,\epsilon^{\mu\alpha\beta\lambda}J_{\alpha\beta}\bar n_\lambda.
\end{equation}
In particular, we have $K^+_n=0$ and $J^-_n=0$.

The LF operators can naturally be expressed in terms of the standard ones in instant form. We can write in general
\begin{equation}
n^\mu=(1,\uvec n)\Lambda,\qquad \bar n^\mu=(1,-\uvec n)/2\Lambda
\end{equation}
with $\uvec n$ a unit vector representing in the Lorentz frame $\ucal S$ the direction of motion of the massless observer\footnote{In practice, one usually makes the canonical choice $\uvec n=-\uvec e_z$ corresponding to the viewpoint of a photon moving along the $-z$ direction.}, and $\Lambda$ some nonzero scaling factor. The LF longitudinal momentum and energy are then given by
\begin{equation}
P^+=(P^0-\uvec P\cdot\uvec n)\Lambda,\qquad P^-=(P^0+\uvec P\cdot\uvec n)/2\Lambda.
\end{equation}
The transverse momentum components $P^i_\perp$ are obviously the same in both instant and front forms. The covariant LF Lorentz generators correspond to particular combinations of covariant instant form boost and rotation generators
\begin{align}
K^\mu_n&=K^\mu_u(u\cdot n)-u^\mu (K_u\cdot n)+\epsilon^{\mu\nu\alpha\beta}n_\nu J_{u\alpha}u_\beta,\\
J^\mu_n&=J^\mu_u(u\cdot\bar n)-u^\mu (J_u\cdot\bar n)-\epsilon^{\mu\nu\alpha\beta}\bar n_\nu K_{u\alpha}u_\beta.
\end{align}
Note that in the $\ucal S_u$ frame, the longitudinal components in both forms are simply proportional to each other
\begin{equation}
K^-_n\overset{\ucal S_u}{=}\uvec K\cdot\uvec n,\qquad J^+_n\overset{\ucal S_u}{=}-\uvec J\cdot\uvec n,
\end{equation}
whereas the transverse components get mixed
\begin{equation}
\uvec K_{n\perp}\overset{\ucal S_u}{=}\left[\uvec K_\perp +(\uvec n\times\uvec J)_\perp\right]\Lambda,\qquad \uvec J_{n\perp}\overset{\ucal S_u}{=}\left[\uvec J_\perp+(\uvec n\times\uvec K)_\perp\right]/2\Lambda.
\end{equation}

\subsection{Center of light-front momentum}

Substituting $u^\mu$ by $n^\mu$ in Eq.~\eqref{dipoledef}, we see that the role of inertia in the LF formalism is taken over by the LF momentum. The coordinates of the LF centroid (or center of LF momentum) are then defined by\footnote{One could also in principle define the center of LF energy $R^\mu_{\bar n}(x^+)=\frac{1}{P^-}\int\ud^3_nx\,x^\mu T^{+-}(x)$, which is less interesting because $P^-$ does not leave the LF hyperplane $x^+=0$ invariant, unlike $P^+$.}
\begin{equation}
R^\mu_n(x^+)=\frac{1}{P^+}\int\ud^3_nx\,x^\mu T^{++}(x).
\end{equation}
Provided that the ISC holds, the LF centroid moves along a straight line with constant LF velocity
\begin{equation}\label{LFCM}
R^\mu_n(x^+)=x^+\,\frac{P^\mu}{P^+}+Y^\mu_n\qquad \text{with}\qquad Y^\mu_n=\frac{J^{\mu +}}{P^+}.
\end{equation}
Looking at the $\mu=0$ component allows us to relate the LF time coordinate with the ordinary time coordinate in $\ucal S$
\begin{equation}
x^0=x^+\,\frac{P^0}{P^+}+Y^0_n
\end{equation}
and to reparametrize the LF centroid worldline as follows
\begin{equation}
X^\mu_n(x^0)\equiv R^\mu_n(x^+(x^0))=\left(x^0-Y^0_n\right)\frac{P^\mu}{P^0}+Y^\mu_n.
\end{equation}
This shows that the LF centroid moves along a line parallel to the instant form centroids with the same constant velocity.

\subsection{M\o ller's circle}

One can naturally also express the LF time coordinate in terms of the proper time $\tau$
\begin{equation}
\tau=x^+\,\frac{M}{P^+}+\frac{Y_n\cdot P}{M}.
\end{equation}
The LF centroid worldline then reads
\begin{equation}
\widetilde X^\mu_n(\tau)\equiv R^\mu_n(x^+(\tau))=\left(\tau-\frac{Y_n\cdot P}{M}\right)\frac{P^\mu}{M}+Y^\mu_n,
\end{equation}
and appears to be shifted relative to the CM by
\begin{equation}\label{covshiftLF}
\widetilde Q^\mu_n\equiv \widetilde X^\mu_n(\tau)-R^\mu_\star(\tau)=-\frac{S^{\mu\nu}_n P_\nu}{M^2}=\frac{S^{\mu+}_\star}{P^+}=-\frac{\epsilon^{\mu\alpha\beta+}S_{\star\alpha} P_\beta}{M P^+}.
\end{equation}
Note that the shift is manifestly independent of the scaling factor $\Lambda$, since both the numerator and the denominator involve the same number of contractions with $n^\mu$. In the CM frame, the shift reduces to
\begin{equation}\label{shiftLF}
\widetilde{\uvec Q}_n\overset{\ucal S_\star}{=}-\frac{\uvec n\times\uvec S_0}{M}.
\end{equation}
Since $\uvec n$ is a unit vector, we see that the set of all LF centroids forms a circle corresponding to the boundary of M\o ller's disk. This is in line with the interpretation of the LF formalism as corresponding to the viewpoint of massless observers, i.e. observers moving at the speed of light.

\subsection{Impact-parameter distributions}

The LF formalism is well suited to study the internal structure of the nucleon, see e.g.~\cite{Brodsky:1997de,Meissner:2009ww,Lorce:2011dv,Lorce:2013pza}. In particular, non-perturbative correlation functions extracted from deeply virtual Compton scattering experiments, known as generalized parton distributions (GPDs), have attracted a lot of attention in the last two decades~\cite{Diehl:2003ny,Belitsky:2005qn,Bacchetta:2016ccz}. The reason for this is because GPDs were shown to give access to the EMT and hence to the AM content of the nucleon~\cite{Ji:1996ek}, and to provide tomographic pictures of the internal structure in impact-parameter space~\cite{Burkardt:2000za,Burkardt:2002hr}. One can even map out the distribution of AM in impact-parameter space~\cite{Lorce:2017wkb}.

It has been observed that the impact-parameter distributions (IPDs) get distorted whenever the state is transversely polarized~\cite{Burkardt:2002hr,Miller:2010nz,Lorce:2015sqe}. These distortions are understood as originating from a relativistic artifact of the LF formalism associated with the internal OAM of quarks and gluons inside the target. When the target is a charged elementary particle, there is no substructure at leading order in QED. The internal OAM must therefore vanish at that order, leading immediately to an unambiguous definition of the natural values for the electromagnetic moments~\cite{Lorce:2009bs}. In particular, we found that the gyromagnetic ratio of elementary particles is given by $g=2$ at tree level for any spin.
\newline

Using the following general parametrization for the matrix elements of the Belinfante-Rosenfeld EMT in a spin-$\frac{1}{2}$ state
\begin{equation}\label{param}
\begin{split}
\langle p',\mathsf S'|T^{\mu\nu}_q(0)|p,\mathsf S\rangle=&\overline u(p',\mathsf S')\left[\frac{P^{\{\mu}\gamma^{\nu\}}}{2}\,A_q(\Delta^2)+\frac{P^{\{\mu}i\sigma^{\nu\}\lambda}\Delta_\lambda}{4M}\,B_q(\Delta^2)\right.\\
&\left.+\frac{\Delta^\mu\Delta^\nu-\eta^{\mu\nu}\Delta^2}{M}\,C_q(\Delta^2)+M\eta^{\mu\nu}\bar C_q(\Delta^2)\right]u(p,\mathsf S),
\end{split}
\end{equation} 
where $a^{\{\mu}b^{\nu\}}=a^\mu b^\nu+a^\nu b^\mu$ and with $P=\frac{p'+p}{2}$ the average momentum, $\Delta=p'-p$ the momentum transfer, $M$ the mass, and $\mathsf S$ ($\mathsf S'$) the initial (final) polarization four-vector of the state, Ji derived in the seminal paper~\cite{Ji:1996ek} a relation between the quark contribution to internal AM and the energy-momentum form factors
\begin{equation}\label{Jirel}
J_q=\tfrac{1}{2}\left[A_q(0)+B_q(0)\right].
\end{equation}
Moreover, he showed that these energy-momentum form factors can be expressed in terms of twist-2 quark vector GPDs
\begin{equation}
A_q(0)=\int\ud x\,x H_q(x,0,0),\qquad B_q(0)=\int\ud x\,x E_q(x,0,0),
\end{equation}
where $x$ is the fraction of LF momentum carried by the quark. For a recent review of the GPD phenomenology, see~\cite{Kumericki:2016ehc}.

Burkardt investigated in Ref.~\cite{Burkardt:2005hp} the relation between AM and transverse distortions of IPDs. Focusing on the ``good'' LF component of the Belinfante-Rosenfeld quark EMT, he found for a nucleon at rest with transverse polarization
\begin{equation}\label{Burkardt}
\langle\psi|\int\ud^3_nx\,x^i_\perp T^{++}_q(x)|\psi\rangle\big|_{x^+=0}\overset{\ucal S_\star}{=}\ucal N\, \epsilon^{ij}_\perp \mathsf S^j_\perp J_q P^+,
\end{equation}
where $\epsilon^{ij}_\perp\equiv\epsilon^{ij+-}$ and $\ucal N$ is a normalization factor depending on the wave packet $|\psi\rangle$. A similar result was obtained using a tower of twist-2 operators generalizing the Belinfante-Rosenfeld AM density~\cite{Hoodbhoy:1998yb} or the Pauli-Luba\'nski pseudo-vector~\cite{Ji:2012sj}. In his derivation, Burkardt insisted on the fact that $|\psi\rangle$ must be a delocalized state centered around the origin. Indeed, if one uses the standard transversely polarized LF state localized at the LF centroid, one would find $\frac{1}{2}B_q(0)$ instead of $J_q$ on the RHS of Eq.~\eqref{Burkardt}. Based on the Melosh-Wigner rotation relating canonical polarizations in instant and front forms~\cite{Wigner:1939cj,Melosh:1974cu,Chung:1988my}, Burkardt argued that the LF wave packet of a transversely polarized spin-$\frac{1}{2}$ state must be shifted sideways by half a Compton wavelength relative to the instant form one. This shift adds a contribution proportional to $\frac{1}{2}A_q(0)$, leading then to Eq.~\eqref{Burkardt}. 

The distortions of the IPDs in the LF formalism and their relation to AM can be more easily understood based on the results obtained in the previous sections. We will work at a fixed LF time $x^+=0$ and drop all references to it in the following for convenience. Let us write the quark LF boost generators as the following telescoping series
\begin{equation}\label{telescoping}
J^{\mu +}_q=\left(R^\mu_{qn}-R^\mu_n\right)P^+_q+\left(R^\mu_n-X^\mu_\star\right)P^+_q+X^\mu_\star P^+_q,
\end{equation}
where $R^\mu_{qn}$, $R^\mu_n$ and $X^\mu_\star$ are the positions at $x^+=0$ of the LF centroid of the quark subsystem, the LF centroid of the whole system, and the CM of the whole system, respectively. The first term of this series can be expressed as
\begin{equation}
\left(R^\mu_{qn}-R^\mu_n\right)P^+_q=\int\ud^3_nx\left(x^\mu-R^\mu_n\right)T^{++}_q(x),
\end{equation}
and represents the LF dipole moment of the quark subsystem relative to the LF centroid. Focusing on the transverse LF components $\mu=i$, we get
\begin{equation}
\left(R^i_{qn\perp}-R^i_{n\perp}\right)P^+_q=\int\ud^2b_\perp\,b^i_\perp\int\ud x^-T^{++}_q(x^-,\uvec b_\perp+\uvec R_{n\perp}),
\end{equation}
where $\uvec b_\perp$ is the usual impact-parameter variable defined as the transverse position relative to the LF centroid. 

The ``position'' of a system is usually identified with that of the canonical reference point\footnote{We used the word position with quotation marks because its definition depends on the observer, and therefore does not represent the actual position of the system which should better be identified with that of the CM, see Section~\ref{elementaryparticle}.}, see Appendix~\ref{AppB}. In the LF formalism, the canonical reference point coincides with the LF centroid. If we set the origin of the coordinate system at the LF centroid, the last two terms in Eq.~\eqref{telescoping} cancel each other and we are left with the first one. This term provides the LF dipole moment associated with the distortions of the IPDs, and reads in terms of the energy-momentum form factors~\cite{Burkardt:2002hr}
\begin{equation}\label{Bcontribution}
\langle\left(R^i_{qn\perp}-R^i_{n\perp}\right)P^+_q\rangle=-\epsilon^{ij}_\perp \mathsf S^j_\perp\,\frac{B_q(0)}{2M}\,P^+.
\end{equation}
In instant form, the canonical reference point coincides in the rest frame with the CM. If we set the origin of the coordinate system at the CM, only the last term in Eq.~\eqref{telescoping} vanishes. Beside the contribution~\eqref{Bcontribution}, we have another one which takes into account the shift between the LF centroid and the CM. Focusing on the transverse LF components, we find
\begin{equation}
\left(R^i_{n\perp}-X^i_{\star\perp}\right)\overset{\ucal S_\star}{=}-\epsilon^{ij}_\perp \mathsf S^j_\perp\,\frac{|\uvec S_0|}{M}
\end{equation}
using Eq.~\eqref{shiftLF}. For a spin-$\tfrac{1}{2}$ state we have $|\uvec S_0|=\tfrac{1}{2}$, giving the shift by half a Compton wavelength between instant and front forms advocated by Burkardt. It simply arises from the dependence of the canonical reference point on the observer. We do not need to invoke Melosh-Wigner rotation effects\footnote{Melosh-Wigner rotation effects and dependence of the canonical reference point on the observer are however related as they both arise from the use of a particular subset of Lorentz transformations in the definition of moving states~\cite{Polyzou:2012ut}. This provides a new perspective on the origin of model relations among various parton dsitributions and OAM~\cite{Lorce:2011zta,Lorce:2011kn}.} and delocalized wave packets. Now since $\langle P^+_q\rangle=A_q(0)P^+$, we finally get
\begin{equation}
\langle\left(R^i_{n\perp}-X^i_{\star\perp}\right)P^+_q\rangle\overset{\ucal S_\star}{=}-\epsilon^{ij}_\perp \mathsf S^j_\perp\,\frac{A_q(0)}{2M}\,P^+.
\end{equation}

Gathering all the contributions, we find that in the rest frame with the CM sitting at the origin of the coordinate system
\begin{equation}\label{Burkardt2}
\langle J^{i+}_q\rangle\overset{\ucal S_\star}{=}-\epsilon^{ij}_\perp\mathsf S^j_\perp J_q\,\frac{P^+}{M},
\end{equation}
which agrees with Eq.~\eqref{Burkardt}. In the RHS, we used the fact that the combination of energy-momentum form factors $\tfrac{1}{2}\left[A_q(0)+B_q(0)\right]$ is precisely the one giving the quark contribution to the nucleon spin $J_q$~\eqref{Jirel}. 

\subsection{Interpretation of Ji's relation}

Twenty years ago, Ji looked for a Lorentz-invariant spin sum rule. For a spin-$\tfrac{1}{2}$ state, he wrote~\cite{Ji:1996kb}
\begin{equation}
\frac{1}{2}=\sum_a\langle p,\mathsf S|\mathsf S\cdot\ucal W_a(0)|p,\mathsf S\rangle,
\end{equation}
where the Pauli-Luba\'nski pseudo-vector $W^\mu=\sum_a \ucal W^\mu_a(x^0)$ has been decomposed into contributions associated with the various types of constituents. Choosing the $z$-axis along the momentum of the system and the longitudinal polarization $\uvec{\mathsf S}\parallel\uvec p$, he obtained the helicity sum rule
\begin{equation}
\frac{1}{2}=\sum_a\langle p,+\tfrac{1}{2}|J^z_a(0)|p,+\tfrac{1}{2}\rangle.
\end{equation}
In particular, the quark contribution to the nucleon spin is defined as
\begin{equation}
J_q=\langle p,+\tfrac{1}{2}|J^z_q(0)|p,+\tfrac{1}{2}\rangle,
\end{equation}
and can be expressed in terms of the energy-momentum form factors as in Eq.~\eqref{Jirel}. This matrix element is valid for any momentum $\uvec p$ as long as one considers longitudinal polarization.
\newline

Reading Eq.~\eqref{Burkardt} backwards suggests another interpretation of Ji's relation~\eqref{Jirel} in terms of transverse polarization instead of longitudinal polarization~\cite{Burkardt:2005hp,Ji:2012sj}. It can be regarded as the sum of two contributions: a term $\frac{1}{2}B_q(0)$ arising from the distortion of the IPD in the LF formalism when the state is transversely polarized, supplemented by a term $\frac{1}{2}A_q(0)$ arising from an overall transverse shift when going from transversely polarized states at rest in instant form to front form. Pushing the interpretation further, it has even been suggested~\cite{Hoodbhoy:1998yb,Burkardt:2005hp,Ji:2012sj} that the GPD combination $J_q(x)\equiv\frac{x}{2}\left[H_q(x,0,0)+E_q(x,0,0)\right]$ should be regarded as the distribution of quark internal AM in $x$-space for a transversely polarized target. This simple partonic interpretation is however not well founded~\cite{Leader:2013jra,Liu:2015xha}, one of the reasons being that although the quantity $J_q$ appears on the RHS of Eq.~\eqref{Burkardt}, the LHS corresponds actually to the matrix element of the transverse LF boost generators at $x^+=0$ and not the transverse AM as one would naively expect~\cite{Leader:2012md,Harindranath:2012wn}. Moreover, we note that Eq.~\eqref{Burkardt} provides a relation between boost generators and spin only at the level of matrix elements. Since interpretations made at the level of matrix elements may be misleading~\cite{Lorce:2017xzd}, one should first determine whether this relation remains valid at the operator level. We doubt this is possible because the dependence on the momentum fraction $x$, being obtained through a non-local operator, cannot unambiguously be related to Ji's AM~\cite{Hatta:2012cs,Lorce:2012ce,Liu:2015xha}.

Ji, Xiong and Yuan~\cite{Ji:2012vj,Ji:2013tva} tried to justify the above alternative interpretation of Ji's relation starting from the quark contribution to the Pauli-Luba\'nski pseudo-vector
\begin{equation}
\mathcal W^\mu_q(x^+)\equiv\tfrac{1}{2}\,\epsilon^{\mu\alpha\beta\lambda}J_{q\alpha\beta}(x^+)P_\lambda.
\end{equation}
To do so, they discarded by hand an annoying term involving the $\bar C_q(0)$ energy-momentum form factor, motivated by the fact that $\sum_a\bar C_a(\Delta^2)=0$ as a consequence of the conservation of the EMT. This is obviously not an acceptable argument when quark and gluon contributions are considered separately. The annoying term has been derived explicitly in both instant and front forms, and shown to depend on the observer~\cite{Leader:2012ar,Hatta:2012jm,Harindranath:2013goa}.

Once again, we can easily understand the above observations based on the results obtained in the previous sections. The quark contribution to the generalized AM tensor at a fixed LF time can be written as
\begin{equation}\label{Jqdec}
J^{\alpha\beta}_q(x^+)=X^\alpha_\star(x^+)P^\beta_q(x^+)-X^\beta_\star(x^+)P^\alpha_q(x^+)+S^{\alpha\beta}_{q,\star}(x^+).
\end{equation}
The quark contribution to the Pauli-Luba\'nski pseudo-vector introduced by Ji, Xiong and Yuan then reads
\begin{equation}
\mathcal W^\mu_q(x^+)=\epsilon^{\mu\alpha\beta\lambda}X_{\star\alpha}(x^+)P_{q\beta}(x^+)P_\lambda+\tfrac{1}{2}\,\epsilon^{\mu\alpha\beta\lambda}S_{q,\star\alpha\beta}(x^+)P_\lambda.
\end{equation}
Because of the first term on the RHS, we see that $\mathcal W^\mu_q(x^+)$ does not represent in general the quark contribution to the nucleon spin despite the fact that $\sum_a\mathcal W^\mu_a(x^+)=W^\mu$. When summed over all quark and gluons contributions, this first term does however vanish owing to energy-momentum conservation $\sum_aP^\mu_a(x^+)=P^\mu$. At the inital LF time $x^+=0$, we get
\begin{equation}
\langle\mathcal W^\mu_q\rangle=\epsilon^{\mu\alpha\beta\lambda}\langle X_{\star\alpha}P_{q\beta}\rangle P_\lambda+\langle J^\mu_q\rangle.
\end{equation}
Treating with care the matrix elements as explained in Appendix~\ref{AppB}, we find using the parametrization~\eqref{param} that the first term on the RHS is precisely the annoying one proportional to $\bar C_q(0)$. Since this term is external, it depends on the choice of origin. Setting as usual the origin at the canonical reference point explains the observer dependence found in~\cite{Leader:2012ar,Hatta:2012jm,Harindranath:2013goa}. Clearly, the only proper ways to get rid of the annoying term is to either restrict ourselves to the longitudinal component like Ji did originally~\cite{Ji:1996kb}, or set the origin at the CM of the system. Note that based on Eq.~\eqref{Jqdec}, one can also easily understand the frame dependence of the transverse AM decomposition obtained in instant form by Leader~\cite{Leader:2011cr,Leader:2013jra}.

\section{Conclusions}

Motivated by the question of orbital angular momentum in hadronic physics, we reviewed the concept of relativistic center of mass in field theory. We extended the discussion to asymmetric energy-momentum tensors and the light-front formalism which constitute the most suitable framework to study the nucleon internal structure.

We found that the canonical reference point with respect to which orbital angular momentum is defined in field theory depends on the observer. This dependence arises because of the quantization of angular momentum in a relativistic theory, and provides a simple explanation for the presence of a $p$-wave in the plane-wave solutions to Dirac equation. It clarifies the
difference between longitudinal and transverse spin sum rules, and the origin of various induced shifts and distortions observed in the distributions defined within the light-front formalism.

The results presented in this work are expected to provide a new perspective on various phenomena like e.g. Thomas precession, Zitterbewegung, and other effects associated with relativistic spin-orbit coupling.

\section*{Acknowledgement}

We are thankful to C. Kopper for his careful reading and comments about the present work. This work has been supported by the Agence Nationale de la Recherche under the project ANR-16-CE31-0019.

\appendix

\section{Poincar\'e algebra}\label{AppA}

The generators $P^\mu$ and $J^{\alpha\beta}$ satisfy the Poincar\'e algebra defined by the following set of (equal-time) Poisson brackets\footnote{Since at the classical level the generators are functionals of the fields, the Poisson brackets of two generators are defined as $\{A,B\}_\text{PB}=\int\ud^3x\sum_a\left[\frac{\delta A}{\delta\phi_a(x)}\frac{\delta B}{\delta\pi_a(x)}-\frac{\delta A}{\delta\pi_a(x)}\frac{\delta B}{\delta\phi_a(x)}\right]$, where $\pi_a(x)$ is the conjugate field of $\phi_a(x)$. In the quantum theory, Poisson brackets are replaced by the standard commutators $\{A,B\}_\text{PB}\mapsto\frac{1}{i}[A,B]$.}~\cite{Greiner:1996zu}
\begin{align}
\{P^\mu,P^\nu\}_\text{PB}&=0,\label{PPbracket}\\
\{P^\mu,J^{\alpha\beta}\}_\text{PB}&=\eta^{\mu\alpha}P^\beta-\eta^{\mu\beta}P^\alpha,\label{PJbracket}\\
\{J^{\mu\nu},J^{\alpha\beta}\}_\text{PB}&=\eta^{\mu\alpha}J^{\beta\nu}-\eta^{\mu\beta}J^{\alpha\nu}+\eta^{\nu\alpha}J^{\mu\beta}-\eta^{\nu\beta}J^{\mu\alpha}.
\end{align}
The first bracket indicates the four-momentum is invariant under spacetime translations. In particular, it is conserved in the sense that it does not depend on time. The other two brackets enforce the components of $P^\mu$ to transform as a Lorentz four-vector and the components of $J^{\alpha\beta}$ to transform an (antisymmetric) rank-2 Lorentz tensor. Contracting Eqs.~\eqref{PPbracket} and~\eqref{PJbracket} with $P_\mu$ shows that $P^2$ has vanishing Poisson bracket with all the Poincar\'e generators, and can therefore be used as a frame-independent label of the system.

The Poincar\'e algebra becomes more transparent when expressed in terms of the generators defined in some Lorentz frame $\ucal S$. Denoting by $H=P^0$ the Hamiltonian of the system, one finds that
\begin{equation}\label{Hbrack}
\{P^i,H\}_\text{PB}=0,\qquad
\{J^i,H\}_\text{PB}=0,\qquad
\{K^i,H\}_\text{PB}=-P^i.
\end{equation}
Since the Poincar\'e generators are time-independent, these relations indicate that only the boost generators involve the time coordinate explicitly owing to
\begin{equation}
0=\dot K^i=\{K^i,H\}_\text{PB}+\partial_tK^i
\end{equation}
with $\dot K^i=\ud K^i/\ud t$. At the same time, the relations~\eqref{Hbrack} indicate that the energy of the system is invariant under translations and rotations, but gets mixed up with momentum under boosts.

One gets also from the Poincar\'e algebra
\begin{equation}
\{P^i,J^j\}_\text{PB}=\epsilon^{ijk}P^k,\qquad
\{J^i,J^j\}_\text{PB}=\epsilon^{ijk}J^k,\qquad
\{K^i,J^j\}_\text{PB}=\epsilon^{ijk}K^k
\end{equation}
that $\uvec P$, $\uvec J$ and $\uvec K$ transform as ordinary three-vectors under rotations. Moreover, these relations tell us that total AM depends explicitly on the coordinates of some particular point owing to
\begin{equation}\label{origindep}
0=\ud J^i/\ud x^j=-\{J^i,P^j\}_\text{PB}+\nabla^j J^i,
\end{equation}
and gets mixed up with boosts under boosts. Total AM contains OAM which is defined with respect to a pivot, and so explicitly depends on the coordinates $\uvec X$ of the latter. In practice, the pivot is identified with the origin of the coordinate system $\uvec X=\uvec 0$, see Eq.~\eqref{convdef}. One therefore omits to write this dependence and identifies $\uvec x$ with the position relative to the pivot. Under an active translation generated by the Poisson brackets, the system is translated by an infinitesimal amount, but the pivot remains at the origin. This changes the total AM and explains the nonvanishing of $\{J^i,P^j\}_\text{PB}$. The latter can however be compensated by an infinitesimal translation of the pivot represented by the term $\nabla^j J^i$. In other words, Eq.~\eqref{origindep} expresses the invariance of total AM when both the system and the pivot are translated.

The last set obtained from the Poincar\'e algebra reads
\begin{equation}
\{P^i,P^j\}_\text{PB}=0,\qquad
\{K^i,P^j\}_\text{PB}=-\delta^{ij}H,\qquad
\{K^i,K^j\}_\text{PB}=-\epsilon^{ijk}J^k.
\end{equation}
These relations indicate that contrary to momentum, boosts do also depend explicitly on the coordinate of some particular point. Moreover, they confirm that boosts mix momentum with energy, and boosts with rotations.

\subsection{Poincar\'e generators relative to a generic frame}

The previous discussion about the Poincar\'e generators can be done in a more covariant way.  Note that the object $u^\mu$ introduced in Section~\ref{KJ} is an auxiliary four-vector, in the sense that it transforms as a four-vector under a change of reference frame $\ucal S$ (passive transformation), but has vanishing Poisson brackets with all Poincar\'e generators (active transformation).

Just like the generators of Lorentz transformations~\eqref{Jdec} can be decomposed into generators of boosts and rotations relative to $\ucal S_u$, the four-momentum can be decomposed into Hamiltonian and momentum operators
\begin{equation}
P^\mu=H_u u^\mu+P^\mu_u,
\end{equation}
where $H_u\equiv P\cdot u$ and $P^\mu_u\equiv \Delta^{\mu\nu}_uP_\nu$ with the projector $\Delta^{\mu\nu}_u\equiv\eta^{\mu\nu}-u^\mu u^\nu$. In terms of these covariant generators, the Poincar\'e algebra reads
\begin{equation}\label{restalg1}
\begin{aligned}
\{H_u,P^\mu_u\}_\text{PB}&=0,&
\{P^\mu_u,P^\nu_u\}_\text{PB}&=0,\\
\{H_u,J^\nu_u\}_\text{PB}&=0,&
\{A^\mu_u,J^\nu_u\}_\text{PB}&=-\epsilon^{\mu\nu\alpha\beta}A_{u\alpha} u_\beta\quad\text{with}\quad A^\mu_u=P^\mu_u,K^\mu_u,J^\mu_u,\\
\{H_u,K^\mu_u\}_\text{PB}&=P^\mu_u,&
\{P^\mu_u,K^\nu_u\}_\text{PB}&=-\Delta^{\mu\nu}_uH_u,\\
\{K^\mu_u,K^\nu_u\}_\text{PB}&=\epsilon^{\mu\nu\alpha\beta}J_{u\alpha} u_\beta.
\end{aligned}
\end{equation}

Among all the possible Lorentz frames $\ucal S_u$, a special role is played by the system rest frame $\ucal S_\star$. This frame is identified by the four-velocity $u^\mu_\star=p^\mu/m$, with $m$ the proper mass and $p^\mu$ the four-momentum obtained from the evaluation of the spacetime translation generators $P^\mu$ for the particular field configuration describing the system\footnote{In the quantum theory, $p^\mu$ corresponds to the expectation value of the four-momentum operator $P^\mu$.}. In particular, we have
\begin{equation}
m\uvec K_\star=p^0\uvec K +(\uvec p\times\uvec J),\qquad m\uvec J_\star=p^0\uvec J -(\uvec p\times\uvec K),
\end{equation}
which are naturally reminiscent of the Lorentz transformation laws for the electric and magnetic fields. Note that because of $\{K^i,P^j\}_\text{PB}=-\delta^{ij}H$, it is possible to set the field evaluation of the boost generators to zero through a suitable translation. Denoting the field evaluation of the generators by the corresponding lower case letters, we get
\begin{equation}
m\uvec k_\star=\uvec p\times\uvec j,\qquad m\uvec j_\star=p^0\uvec j.
\end{equation}
We therefore find that $\uvec A\equiv m\uvec k_\star$ is nothing but the (relativistic version of the) Laplace-Runge-Lenz vector of Classical Mechanics in absence of external forces.

\subsection{Poincar\'e generators relative to the instantaneous rest frame}

Instead of working with an auxiliary four-vector $u^\mu$, we can use the generators of spacetime translations $P^\mu$ to define
\begin{equation}
N^\mu\equiv -J^{\mu\nu}P_\nu,\qquad W^\mu\equiv\tfrac{1}{2}\,\epsilon^{\mu\alpha\beta\lambda}J_{\alpha\beta}P_\lambda,
\end{equation}
where $W^\mu$ is the standard Pauli-Luba\'nski pseudo-vector, and therefore write
\begin{equation}
J^{\alpha\beta}P^2=-N^\alpha P^\beta+N^\beta P^\alpha-\epsilon^{\alpha\beta\mu\nu}W_\mu P_\nu.
\end{equation}
Clearly, $N^\mu$ and $W^\mu$ are both orthogonal to $P^\mu$.

Unlike $K^\mu_\star=-J^{\mu\nu}p_\nu/m$ and $J^\mu_\star=\tfrac{1}{2}\,\epsilon^{\mu\alpha\beta\lambda}J_{\alpha\beta}p_\lambda/m$, the objects $N^\mu$ and $W^\mu$ behave as four-vectors under active Lorentz transformations generated by the Poisson brackets
\begin{equation}\label{LFVT}
\{A^\mu,J^{\alpha\beta}\}_\text{PB}=\eta^{\mu\alpha}A^\beta-\eta^{\nu\beta}A^\alpha\qquad\text{with}\qquad A^\mu=N^\mu,W^\mu.
\end{equation}
They satisfy
\begin{equation}\label{restalg2}
\begin{aligned}
\{W^\mu,P^\nu\}_\text{PB}&=0,&\quad \{W^\mu,W^\nu\}_\text{PB}&=-\epsilon^{\mu\nu\alpha\beta}W_\alpha P_\beta,\\
\{N^\mu,P^\nu\}_\text{PB}&=\eta^{\mu\nu}M^2-P^\mu P^\nu,& \{W^\mu,N^\nu\}_\text{PB}&= W^\nu P^\mu,\\
\{N^\mu,N^\nu\}_\text{PB}&=J^{\mu\nu}M^2.
\end{aligned}
\end{equation}
In particular, it is readily seen that $W^2$ (like $P^2$) has vanishing Poisson brackets with all the Poincar\'e generators. The first two relations are simple to interpret. They indicate that the relativistic spin $S^\mu=W^\mu/\sqrt{P^2}$ is independent of the choice of origin, and obeys standard commutation relations for AM only in the rest frame.

The difference between the algebras~\eqref{restalg1} and~\eqref{restalg2} appears in the Poisson brackets $\{W^\mu,N^\nu\}_\text{PB}\not\propto \{J^\mu_\star,K^\nu_\star\}_\text{PB}$ and $\{N^\mu,N^\nu\}_\text{PB}\not\propto \{K^\mu_\star,K^\nu_\star\}_\text{PB}$. This can be understood by the fact that $K^\mu_\star$ and $J^\mu_\star$ represent the boost and rotation generators relative to the rest frame of the system determined \emph{prior} any active Poincar\'e transformation, whereas $N^\mu/\sqrt{P^2}$ and $W^\mu/\sqrt{P^2}$ represent the boost and rotation generators relative to the rest frame of the system at the moment of their action. Since active boosts change the frame in which the system is at rest, subsequent $N^\mu$ and $W^\mu$ do not coincide anymore with $mK^\mu_\star$ and $mJ^\mu_\star$.

\subsection{Algebra of the covariant position, orbital and spin angular momentum}

In Section~\ref{sec3}, we found that the position of the center of inertia, and the associated external and internal parts of AM are constructed from the Poincar\'e generators as follows
\begin{equation}
\uvec R(x^0)=(x^0\uvec P-\uvec K)/H,\qquad \uvec L=-(\uvec K\times\uvec P)/H,\qquad \uvec S=\uvec J+(\uvec K\times\uvec P)/H.
\end{equation}
They all transform as ordinary three-vectors under rotations
\begin{equation}
\{A^i,J^j\}_\text{PB}=\epsilon^{ijk}A^k\qquad\text{with}\qquad A^i=R^i,L^i,S^i,
\end{equation}
and satisfy familiar Poisson brackets with energy
\begin{equation}
\{R^i,H\}_\text{PB}=P^i/H,\qquad\{L^i,H\}_\text{PB}=0,\qquad\{S^i,H\}_\text{PB}=0,
\end{equation}
and momentum
\begin{equation}
\{R^i,P^j\}_\text{PB}=\delta^{ij},\qquad\{L^i,P^j\}_\text{PB}=\epsilon^{ijk}P^k,\qquad\{S^i,P^j\}_\text{PB}=0.
\end{equation}
The remaining Poisson brackets~\cite{Born:1935ap}
\begin{equation}
\begin{aligned}
\{R^i,R^j\}_\text{PB}&=-\epsilon^{ijk}S^k/H^2,\\
\{R^i,L^j\}_\text{PB}&=\epsilon^{ijk}R^k-(\delta^{ij}\delta^{kl}-\delta^{ik}\delta^{jl})P^kS^l/H^2,\\
\{R^i,S^j\}_\text{PB}&=(\delta^{ij}\delta^{kl}-\delta^{ik}\delta^{jl})P^kS^l/H^2,\\
\{L^i,L^j\}_\text{PB}&=\epsilon^{ijk}L^k-\epsilon^{ijk}P^k(\uvec P\cdot\uvec S)/H^2,\\
\{L^i,S^j\}_\text{PB}&=\epsilon^{ijk}P^k(\uvec P\cdot\uvec S)/H^2,\\
\{S^i,S^j\}_\text{PB}&=\epsilon^{ijk}S^k-\epsilon^{ijk}P^k(\uvec P\cdot\uvec S)/H^2,
\end{aligned}
\end{equation}
differ from the familiar ones by terms proportional to $S^i/H^2$. In the nonrelativistic limit, the contribution of these terms vanishes as one can easily see by reinstating the factors of $c$ in the above expressions. They can therefore be understood as relativistic corrections. Note also that in the system rest frame, all the relativistic corrections disappear except for $\{R^i,R^j\}_\text{PB}$.

For the corresponding covariant quantities defined by an observer in $\ucal S_u$, we have
\begin{equation}
\begin{aligned}
X^\mu_u(x^0)=(x^0P^\mu-&K^\mu_u)/H_u,\qquad
L^\mu_u=-\epsilon^{\mu\alpha\beta\lambda}K_{u\alpha}P_{u\beta} u_\lambda/H_u,\\
 S^\mu_u&=J^\mu_u+\epsilon^{\mu\alpha\beta\lambda}K_{u\alpha}P_{u\beta} u_\lambda/H_u.
\end{aligned}
\end{equation}
They satisfy the following Poisson brackets with the covariant energy
\begin{equation}
\{X^\mu_u,H_u\}_\text{PB}=P^\mu_u/H_u,\qquad \{L^\mu_u,H_u\}_\text{PB}=0,\qquad \{S^\mu_u,H_u\}_\text{PB}=0,
\end{equation}
and the covariant momentum
\begin{equation}
\{X^\mu_u,P^\nu_u\}_\text{PB}=-\Delta^{\mu\nu}_u,\qquad \{L^\mu_u,P^\nu_u\}_\text{PB}=-\epsilon^{\mu\nu\alpha\beta}P_{u\alpha} u_\beta,\qquad \{S^\mu_u,P^\nu_u\}_\text{PB}=0.
\end{equation}
The remaining Poisson brackets read
\begin{equation}
\begin{aligned}
\{X^\mu_u,X^\nu_u\}_\text{PB}&=\epsilon^{\mu\nu\alpha\beta}S_{u\alpha} u_\beta/H^2_u,\\
\{X^\mu_u,L^\nu_u\}_\text{PB}&=-\epsilon^{\mu\nu\alpha\beta}X_{u\alpha} u_\beta-(\Delta^{\mu\nu}_u\Delta^{\alpha\beta}_u-\Delta^{\mu\alpha}_u\Delta^{\nu\beta}_u)P_{u\alpha}S_{u\beta}/H^2_u,\\
\{X^\mu_u,S^\nu_u\}_\text{PB}&=(\Delta^{\mu\nu}_u\Delta^{\alpha\beta}_u-\Delta^{\mu\alpha}_u\Delta^{\nu\beta}_u)P_{u\alpha}S_{u\beta}/H^2_u,\\
\{L^\mu_u,L^\nu_u\}_\text{PB}&=-\epsilon^{\mu\nu\alpha\beta}L_{u\alpha} u_\beta-\epsilon^{\mu\nu\alpha\beta}P_{u\alpha}u_\beta(P_u\cdot S_u)/H^2_u,\\
\{L^\mu_u,S^\nu_u\}_\text{PB}&=\epsilon^{\mu\nu\alpha\beta}P_{u\alpha}u_\beta(P_u\cdot S_u)/H^2_u,\\
\{S^\mu_u,S^\nu_u\}_\text{PB}&=-\epsilon^{\mu\nu\alpha\beta}S_{u\alpha} u_\beta-\epsilon^{\mu\nu\alpha\beta}P_{u\alpha}u_\beta(P_u\cdot S_u)/H^2_u.
\end{aligned}
\end{equation}

For the instantaneous CM operator $R^\mu(\tau)=\tau P^\mu/M-N^\mu/M^2$, we find
\begin{equation}
\begin{aligned}
\{R^\mu,P^\nu\}_\text{PB}&=-\eta^{\mu\nu}+P^\mu P^\nu/M^2,\\
\{R^\mu,R^\nu\}_\text{PB}=J^{\mu\nu}/M^2,&\qquad \{R^\mu,W^\nu\}_\text{PB}=W^\mu P^\nu/M^2.
\end{aligned}
\end{equation}
It follows from Eq.~\eqref{LFVT} that $R^\mu(\tau)$ transforms as a Lorentz four-vector
\begin{equation}
\{R^\mu,J^{\alpha\beta}\}_\text{PB}=\eta^{\mu\alpha}R^\beta-\eta^{\mu\beta}R^\alpha.
\end{equation}

\subsection{Light-front form}

Using the notations $\ucal B_L=J^{+-}$, $\ucal B^i_\perp=J^{+i}$, $\ucal J_L=\frac{1}{2}\,\epsilon^{ij}_\perp J^{ij}$ and $\ucal J^i_\perp=\epsilon^{ij}_\perp J^{-j}$ for the LF boost and rotation generators introduced in Section~\ref{LFcomponent}, the Poincar\'e algebra reads
\begin{align}
\{P^i_\perp,P^+\}_\text{PB}&=0,&\{\ucal B^i_\perp,P^+\}_\text{PB}&=0,&\{\ucal J_L,P^+\}_\text{PB}&=0,\label{Gali}\\
\{P^i_\perp,P^j_\perp\}_\text{PB}&=0,&\{P^i_\perp,\ucal B^j_\perp\}_\text{PB}&=\delta^{ij}_\perp P^+,&\{\ucal B^i_\perp,\ucal B^j_\perp\}_\text{PB}&=0, \\
\{\ucal J_L,P^i_\perp\}_\text{PB}&=\epsilon^{ij}_\perp P^j_\perp,&\{\ucal J_L,\ucal B^i_\perp\}_\text{PB}&=\epsilon^{ij}_\perp \ucal B^j_\perp,\label{Galf}
\end{align}
and
\begin{align}
\{P^i_\perp,P^-\}_\text{PB}&=0,&\{\ucal B^i_\perp,P^-\}_\text{PB}&=-P^i_\perp,&\{\ucal J_L,P^-\}_\text{PB}&=0,\\
\{P^i_\perp,\ucal J^j_\perp\}_\text{PB}&=\epsilon^{ij}_\perp P^-,&\{\ucal J^i_\perp,\ucal J^j_\perp\}_\text{PB}&=0,&\{\ucal B_L,P^\pm\}_\text{PB}&=\pm P^\pm, \\
\{P^+,P^-\}_\text{PB}&=0,&\{P^+,\ucal J^i_\perp\}_\text{PB}&=\epsilon^{ij}_\perp P^j_\perp,&\{P^-,\ucal J^i_\perp\}_\text{PB}&=0,\\
\{\ucal B_L,P^i_\perp\}_\text{PB}&=0,&\{\ucal B_L,\ucal B^i_\perp\}_\text{PB}&=\ucal B^i_\perp,&\{\ucal B_L,\ucal J_L\}_\text{PB}&=0,\\
\{\ucal B^i_\perp,\ucal J^j_\perp\}_\text{PB}&=-\delta^{ij}_\perp \ucal J_L+\epsilon^{ij}_\perp \ucal B_L,&\{\ucal J_L,\ucal J^i_\perp\}_\text{PB}&=\epsilon^{ij}_\perp \ucal J^j_\perp,&\{\ucal B_L,\ucal J^i_\perp\}_\text{PB}&=-\ucal J^i_\perp.
\end{align}
We see that the first set forms a two-dimensional Galilean subgroup where the LF momentum plays the role of a non-relativistic ``mass'' in the transverse plane~\cite{Susskind:1967rg,Susskind:1968zz,Kogut:1969xa}.

In terms of the transverse LF position $\uvec R_\perp=(x^+ \uvec P_\perp-\uvec{\ucal B}_\perp)/P^+$, we have
\begin{align}
\{R^i_\perp,P^+\}_\text{PB}&=0,& \{R^i_\perp,P^j_\perp\}_\text{PB}&=\delta^{ij}_\perp,& \{R^i_\perp,\ucal B^j_\perp\}_\text{PB}&=x^+\delta^{ij}_\perp,\\
\{\ucal J_L,R^i_\perp\}_\text{PB}&=\epsilon^{ij}_\perp R^j_\perp,& \{R^i_\perp,R^j_\perp\}_\text{PB}&=0,
\end{align}
and for the longitudinal LF position $R^-=(x^+ P^--\ucal B_L)/P^+$, we have
\begin{align}
\{R^-,P^+\}_\text{PB}&=-1,& \{R^-,P^i_\perp\}_\text{PB}&=0,& \{R^-,\ucal B^i_\perp\}_\text{PB}&=R^i_\perp,\\
\{\ucal J_L,R^-\}_\text{PB}&=0,& \{R^-,R^i_\perp\}_\text{PB}&=0.
\end{align}
The other Poisson brackets read
\begin{align}
\{R^i_\perp,P^-\}_\text{PB}&=\frac{P^i_\perp}{P^+},& \{\ucal B_L,R^i_\perp\}_\text{PB}&=-x^+\frac{P^i_\perp}{P^+},\\
\{R^i_\perp,\ucal J^j_\perp\}_\text{PB}&=-R^i_\perp\epsilon^{jk}_\perp\frac{P^k_\perp}{P^+}+\delta^{ij}_\perp\frac{\ucal J_L}{P^+}+\epsilon^{ij}_\perp R^-,\\
\{R^-,P^-\}_\text{PB}&=\frac{P^-}{P^+},& \{\ucal B_L,R^-\}_\text{PB}&=-x^+\frac{P^-}{P^+}-R^-,\\
\{R^-,\ucal J^i_\perp\}_\text{PB}&=-R^-\epsilon^{ij}_\perp\frac{P^j_\perp}{P^+}+\frac{\ucal J^i_\perp}{P^+}.
\end{align}

\section{Canonical reference point}\label{AppB}

When it comes to evaluating matrix elements of OAM in a quantum theory, one has to be careful with the treatment of the position variable~\cite{Bakker:2004ib,Leader:2013jra}. The problem is that OAM requires the knowledge of both position and momentum, explaining why a standard plane-wave approach fails. One solution is to consider wave packets~\cite{Bakker:2004ib}, but the price to pay is that calculations usually become quite lengthy and cumbersome. In particular, one has to identify and remove the part associated with the structure of the wave packet, which we are not interested in. An equivalent and much simpler solution is to use the Wigner-Weyl representation~\cite{Case:2008}.

Consider the matrix element of an operator $O$ in some state $|\psi\rangle$ normalized as $\langle\psi|\psi\rangle=1$. It can conveniently be written in the form
\begin{equation}
\langle O\rangle_\psi=\langle \psi|O|\psi\rangle=\uTr[O\rho_\psi]
\end{equation}
with the density operator $\rho_\psi=|\psi\rangle\langle\psi|$. A closed system with mass $M$, average position $\uvec{\ucal R}(x^0)$ and momentum $\uvec P$ defined in the Wigner sense at some time $\ucal R^0(x^0)=x^0$ is represented by the following relativistic phase-space operator\footnote{Note that our normalization factor $2\sqrt{p^0p'^0}$ is consistent with other works. When one neglects relativistic recoil corrections like in~\cite{Ji:2003ak,Belitsky:2003nz}, it reduces to $2M$. If one works in the Breit frame $\uvec P=\uvec 0$ like in~\cite{Polyakov:2002yz}, it reduces to $2P^0$.}
\begin{equation}\label{rhoRP}
\rho_{\uvec{\ucal R},\uvec P}(x^0)\equiv\int\frac{\ud^3\Delta}{(2\pi)^3\,2\sqrt{p^0p'^0}}\,e^{-i\Delta\cdot\ucal R}\,|P-\tfrac{\Delta}{2}\rangle\langle P+\tfrac{\Delta}{2}|,
\end{equation}
where the four-vectors $P=\frac{p'+p}{2}$ and $\Delta=p'-p$ satisfy the constraints $P\cdot\Delta=0$ and $P^2+\frac{\Delta^2}{4}=M^2$ which arise from the mass shell conditions. A similar operator can be introduced within the LF formalism~\cite{Lorce:2011kd,Lorce:2011ni}. 

The phase-space operator is properly normalized $\uTr[\rho_{\uvec{\ucal R},\uvec P}]=1$, and plane waves are recovered by averaging over $\uvec{\ucal R}$
\begin{equation}\label{meanP}
\int\frac{\ud^3\ucal R}{(2\pi)^3\delta^{(3)}(\uvec 0)}\,\rho_{\uvec{\ucal R},\uvec P}(x^0)=\frac{|P\rangle\langle P|}{\langle P|P\rangle}.
\end{equation}
Note that the time dependence drops out because the integral over $\uvec{\ucal R}$ imposes $\Delta=0$. Defining ``position'' states at some time $r^0(x^0)=x^0$ as
\begin{equation}
|r\rangle\equiv\int\frac{\ud^3p}{(2\pi)^3\,\sqrt{2p^0}}\,e^{i p\cdot r}\,|p\rangle
\end{equation}
with normalization $\langle r'|r\rangle=\delta^{(3)}(\uvec r'-\uvec r)$, the phase-space operator~\eqref{rhoRP} can alternatively be expressed as
\begin{equation}
\rho_{\uvec{\ucal R},\uvec P}(x^0)=\int\ud^3Z\,e^{-iP\cdot Z}\,|\ucal R+\tfrac{Z}{2}\rangle\langle \ucal R-\tfrac{Z}{2}|
\end{equation}
with $Z^0=0$. If we average over $\uvec P$, we recover the ``spatial'' density operator
\begin{equation}\label{meanR}
\int\frac{\ud^3P}{(2\pi)^3\delta^{(3)}(\uvec 0)}\,\rho_{\uvec{\ucal R},\uvec P}(x^0)=\frac{|\ucal R\rangle\langle \ucal R|}{\langle\ucal R|\ucal R\rangle}.
\end{equation}

Let us now consider the matrix elements of the OAM operator $J^i_\text{orb}=\epsilon^{ijk}\int\ud^3x\,x^j\,T^{0k}(x)$. After some standard manipulations~\cite{Leader:2013jra}, we find
\begin{align}
\langle J^i_\text{orb}\rangle_{\uvec{\ucal R},\uvec P}(x^0)&=\epsilon^{ijk}\left\{-i\,\frac{\partial}{\partial\Delta^j}\!\left[e^{i\uvec\Delta\cdot\uvec{\ucal R}}\,\frac{\langle P+\tfrac{\Delta}{2}|T^{0k}(0)|P-\tfrac{\Delta}{2}\rangle}{2\sqrt{p^0p'^0}}\right]\right\}_{\Delta=0}\nonumber\\
&=\epsilon^{ijk}\ucal R^j(x^0) P^k+\frac{\epsilon^{ijk}}{2E}\left[-i\,\frac{\partial}{\partial\Delta^j}\langle P+\tfrac{\Delta}{2}|T^{0k}(0)|P-\tfrac{\Delta}{2}\rangle\right]_{\Delta=0},\label{genexpr}
\end{align}
where $E=P^0|_{\Delta=0}=\sqrt{\uvec P^2+M^2}$. These two terms represent the external and internal parts of the OAM defined by the canonical reference point $\uvec{\ucal R}(x^0)$. In practice, one usually works at a fixed initial time $x^0=0$ and sets the origin at the initial average position of the system $\uvec{\ucal R}(0)=\uvec 0$.

In order to determine where this canonical reference point is situated, we consider the matrix elements of the operator $R^i(x^0)=\frac{1}{P^0}\int\ud^3x\,x^i\,T^{00}(x)$ giving the position of the center of inertia. After some algebra, we find
\begin{equation}
\langle R^i\rangle_{\uvec{\ucal R},\uvec P}(x^0)=\ucal R^i(x^0)+\frac{1}{2E^2}\left[-i\,\frac{\partial}{\partial\Delta^i}\langle P+\tfrac{\Delta}{2}|T^{00}(0)|P-\tfrac{\Delta}{2}\rangle\right]_{\Delta=0}.
\end{equation}
For a closed spin-$\frac{1}{2}$ system, we can use the parametrization~\eqref{param} with $A(0)=1$ and $B(0)=\bar C(0)=0$ which arise from the conservation of total linear and angular momenta~\cite{Leader:2013jra}. We then get
\begin{align}
\langle R^i\rangle_{\uvec{\ucal R},\uvec P}(x^0)-\ucal R^j(x^0)&=\frac{1}{2E}\left\{-i\,\frac{\partial}{\partial\Delta^j}\!\left[\overline u(p',\mathsf S')\gamma^0 u(p,\mathsf S)\right]\right\}_{\Delta=0}\nonumber\\
&=\frac{\epsilon^{ijk} P^j\mathsf S^k}{2E(E+M)},
\end{align}
where we used the generic expression for the Dirac bilinear derived in~\cite{Lorce:2017isp}. The shift between the center of inertia and the CM being given by Eq.~\eqref{covshift}
\begin{equation}
\langle R^i\rangle_{\uvec{\ucal R},\uvec P}(x^0)-\langle X^i_\star\rangle_{\uvec{\ucal R},\uvec P}(x^0)=\frac{\epsilon^{ijk} P^j\mathsf S^k}{2ME},
\end{equation}
we conclude that the canonical reference point $\uvec{\ucal R}$ is situated on the segment joining the center of inertia to the CM. In the Breit frame $\uvec P=\uvec 0$, which corresponds to the average CM frame, the canonical reference point coincides with both the center of inertia and the CM. In the infinite-momentum frame $|\uvec P|\to\infty$, the canonical reference point coincides with the center of inertia and is half a Compton wavelength away from the CM. Between these two limiting frames, all three points differ.

Within the LF formalism, one usually works in the symmetric frame $\uvec P_\perp=\uvec 0_\perp$. Using again the results of~\cite{Lorce:2017isp}, we find that the canonical reference point will always coincide with the center of LF momentum in the transverse plane
\begin{align}
\langle R^i_\perp\rangle_{\uvec{\ucal R},\uvec P}(x^+)-\ucal R^i_\perp(x^+)&=\frac{1}{2P^+}\left\{-i\,\frac{\partial}{\partial\Delta^i_\perp}\!\left[\overline u_\text{LF}(p',\mathsf S')\gamma^+ u_\text{LF}(p,\mathsf S)\right]\right\}_{\Delta=0}\nonumber\\
&=0,
\end{align}
in agreement with the discussion in~\cite{Burkardt:2002hr} based on the Galilean subgroup~\eqref{Gali}-\eqref{Galf}.

\end{document}